%% file: conf_sum2008.tex
\documentclass[11pt]{article}
\usepackage{graphicx}
\usepackage{setspace}

\newcommand{\BABARPubYear}    {08}

\newcommand{\BABARConfNumber} {012}
\newcommand{\SLACPubNumber} {13347}

\input babarsym

\long\def\inst#1{\par\nobreak\kern 4pt\nobreak
    {\it #1}\par\vskip 10pt plus 3pt minus 3pt}

\begin{document}
{\pagestyle{empty}

\begin{flushright}
\babar-CONF-\BABARPubYear/\BABARConfNumber \\
SLAC-PUB-\SLACPubNumber \\
July 2008 \\
\end{flushright}

\par\vskip 5cm

\begin{center}
\Large \bf Measurement of the Branching Fractions of the
Color-Suppressed Decays $\Bzb\ra\Dz\piz$, $\Dstarz\piz$,
$\Dz\eta$, $\Dstarz\eta$, $\Dz\omega$, $\Dstarz\omega$,
$\Dz\etapr$, and $\Dstarz\etapr$
\end{center}
\bigskip

\begin{center}
\large The \babar\ Collaboration\\
\mbox{ }\\
\today
\end{center}
\bigskip \bigskip

\begin{center}
\large We report results on the branching fraction ($\BF$)
measurement of the color-suppressed decays $\Bzb\ra\Dz\piz$,
$\Dstarz\piz$, $\Dz\eta$, $\Dstarz\eta$, $\Dz\omega$,
$\Dstarz\omega$, $\Dz\etapr$, and $\Dstarz\etapr$. We measure the
branching fractions $\BF(\Dz\piz)=(2.78\pm 0.08 \pm 0.20)\times
10^{-4}$, $\BF(\Dstarz\piz)=(1.78\pm 0.13\pm 0.23 )\times
10^{-4}$, $\BF(\Dz\eta)=(2.41\pm 0.09  \pm 0.17 )\times 10^{-4}$,
$\BF(\Dstarz\eta)=(2.32\pm 0.13  \pm 0.22 )\times 10^{-4}$,
$\BF(\Dz\omega)=(2.77\pm 0.13  \pm 0.22 )\times 10^{-4}$,
$\BF(\Dstarz\omega)=(4.44\pm 0.23 \pm 0.61 )\times 10^{-4}$,
$\BF(\Dz\etapr)=(1.38\pm 0.12 \pm 0.22 )\times 10^{-4}$ and
$\BF(\Dstarz\etapr)=(1.29\pm 0.23 \pm 0.23 )\times 10^{-4}$, where
the first uncertainty is statistical and the second is systematic.
The result is based on a sample of $(454\pm 5)\times 10^{6} \BB$
pairs collected at the $\Upsilon(4S)$ resonance from 1999 to 2007,
with the $\babar$ detector at the \pep2 storage rings at the
Stanford Linear Accelerator Center. The measurements are compared
to theoretical predictions by factorization, SCET and pQCD. The
presence of final state interactions is confirmed and the
measurements seem to be more in favor of SCET compared to pQCD.
\end{center}

\vfill
\begin{center}

Submitted to the 34$^{\rm th}$ International Conference on High-Energy Physics, ICHEP 08,\\
29 July---5 August 2008, Philadelphia, Pennsylvania.

\end{center}

\vspace{1.0cm}
\begin{center}
{\em Stanford Linear Accelerator Center, Stanford University,
Stanford, CA 94309} \\ \vspace{0.1cm}\hrule\vspace{0.1cm}
Work supported in part by Department of Energy contract DE-AC02-76SF00515.
\end{center}

\newpage
} 

%
%
\input authors_ICHEP2008.tex

\section{INTRODUCTION}
\label{sec:Introduction} Weak decays of hadrons provide a straight
access to the parameters of the CKM matrix and thus to the study
of the \CP violation. Gluon scattering in the final state (Final
State Interactions, or FSI) can modify the decay
dynamics and so must be well understood. The two-body hadronic
decays with a charmed final state, $B\ra Dh$, are of great help in
studying strong-interaction physics related with the confinement
of quarks and gluons into hadrons.

\begin{figure}[h]
\begin{center}
\includegraphics[width=0.500\linewidth]{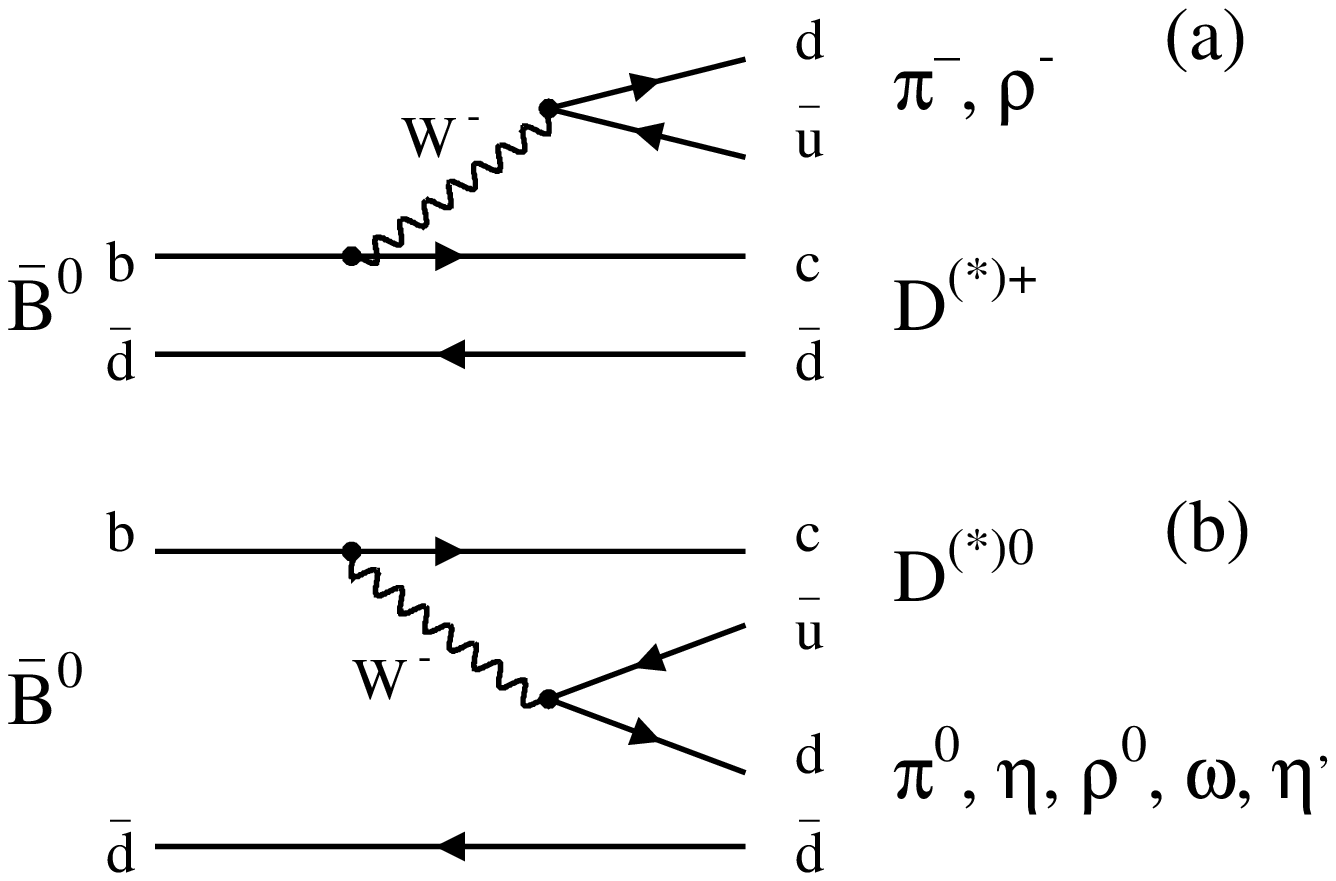}
 \caption{External (a) and internal (b) tree diagrams for $\Bzb\ra Dh$ decays.}
 \label{fig:feynman}
\end{center}
\end{figure}

The decays $B\ra Dh$, where $h$ is a light meson, can proceed
through the emission of a $W^{\pm}$ boson following three possible
diagrams: external, internal (see Fig. \ref{fig:feynman}) or by a
$W^{\pm}$ boson exchange whose contribution is
negligible~\cite{ref:exchangeW}.

In the case of the decays $\Bzb\ra\Dstze\hz$, the major
contribution comes from the internal diagram~\cite{ref:Neubert}. Since mesons are
color single objects, in internal diagrams $\Bzb\ra\Dstze\hz$ the
quarks from the $W^{\pm}$ decay are constrained to have the
anti-color of the spectator quark, which induces a suppression of
internal diagrams in comparison with external ones. Internal
diagrams are so called {\it color-suppressed} and external ones
are called {\it color-favored}.

In the factorization model~
\cite{ref:Neubert,ref:BauerStechWirbel,ref:NeubertPetrov,ref:Deandrea_0},
the non-factorizable interactions in the final state by soft
gluons are neglected. The matrix element in the effective weak
Hamiltonian of the decay $B\ra Dh$ is then factorized into a
product of asymptotic states. Factorization appears to be
successful in the description of the color-favored
decays~\cite{ref:HSW}.

The decays $\Bzb\ra\Dstze\piz$ were first observed by
CLEO~\cite{ref:Cleo2002} and Belle~\cite{ref:Belle2002} with
respectively 9.67 and 23.1$\times 10^6~\BB$ pairs. The Belle
collaboration has also observed the decays $\Dz\eta$ and
$\Dz\omega$ and put upper limits on the $\BF$ of $\Dstarz\eta$ and
$\Dstarz\omega$~\cite{ref:Belle2002}.

The branching fraction ($\BF$) of the color-suppressed decays
$\Bzb\ra\Dstze\piz$, $\Dstze\eta$, $\Dstze\omega$, and $\Dz\etapr$
were measured recently by $\babar$~\cite{ref:Babar2004} with
88$\times 10^6~\BB$ pairs and an upper limit was set on the $\BF$ of
$\Dstarz\etapr$. The Belle collaboration measured with 152$\times
10^6~\BB$ pairs the $\BF$ of $\Bzb\ra\Dstze\hz$, $\hz=\piz$, $\eta$,
$\omega$, and $\etapr$~\cite{ref:Belle2005,ref:Belle2006} and
studied the decays $\Bzb\ra\Dstze\rho^0$ with 388$\times
10^6~\BB$ pairs~\cite{ref:Belle2007}. Many of these measurements showed
a significant disagreement with predictions by
factorization~\cite{ref:Chua}, but stronger experimental
constraints are needed to distinguish between the different models of the
color-suppressed dynamics like
pQCD ({\it perturbative QCD})~\cite{ref:pQCD_1,ref:pQCD_2} or
SCET ({\it Soft Collinear Effective Theory})~\cite{ref:SCET_1,ref:SCET_2,ref:SCET_3}. This
paper reports the branching fraction measurement of eight
color-suppressed decays $\Bzb\ra\Dstze\piz$, $\Dstze\eta$,
$\Dstze\omega$ and $\Dstze\etapr$ with 454$\times 10^6~\BB$ pairs.

\section{THE \babar\ DETECTOR AND DATASET}
\label{sec:babar}

The data used in this analysis were collected with the $\babar$
detector at the \pep2 asymmetric $\epem$ storage ring. The
$\babar$ detector is described in detail in
Ref.~\cite{ref:detector}. Charged particle tracks are
reconstructed using a five-layer silicon vertex tracker (SVT) and
a 40-layer drift chamber (DCH) immersed in a 1.5~T magnetic field.
Tracks are identified as pions or kaons (particle identification
or PID) based on likelihoods constructed from energy loss
measurements in the SVT and the DCH and from Cherenkov radiation
angles measured in the detector of Cherenkov light (DIRC).
Photons are reconstructed from showers measured in the
electromagnetic calorimeter (EMC). Muon and neutral hadron
identification are performed with the instrumented flux return
(IFR).

The data sample consists of an integrated luminosity of
413~$\invfb$ recorded at the $\Upsilon(4S)$ resonance with a
center-of-mass (CM) energy of 10.58~$\gev$, corresponding to
$(454\pm 5)\times 10^6~ \BB$ pairs. A data sample of 41~$\invfb$ with
a CM energy 40~$\mev$ below the $\Upsilon(4S)$ resonance is used
to study background contributions from continuum events
$\epem\ra\qqbar$ ($q=u$, $d$, $s$, $c$).

Samples of simulated Monte Carlo (MC) events were used to
determine signal and background characteristics, optimize
selection criteria and evaluate efficiencies. Simulated events
$\epem\ra\Upsilon(4S)\ra\BpBm,~\Bz\Bzb$, $\epem\ra\qqbar$ ($q=u$,
$d$, $s$) and $\epem\ra\ccbar$ are generated with {\tt EvtGen}~\cite{ref:evtgen}, which interfaces to {\tt
Pythia}~\cite{ref:pythia} and {\tt Jetset}~\cite{ref:jetset}.
Separate samples of exclusive $\Bzb\ra\Dstze\hz$ decays were
generated to evaluate the signal features and the efficiency of
selections on signal. A sample of exclusive $\Bm\ra\Dstze\rho^-$
was generated for the study of that background. All MC samples
include simulation of the $\babar$ detector response generated
through {\tt GEANT4}~\cite{ref:GEANT4}. The integrated luminosity
of the MC samples is about three times the data luminosity for
$\BB$, one times the data luminosity for $\epem\ra\qqbar$ ($q=u$, $d$, $s$) and two times
for $\epem\ra\ccbar$. The equivalent integrated luminosities of
the exclusive simulations range from 50 to 2500 times the data
luminosity.

\section{ANALYSIS METHOD}
\label{sec:Analysis}

\subsection{Event reconstruction}

Charged particles tracks are reconstructed from measurements in
the SVT and$/$or the DCH, and an identification is assigned by the PID
algorithm. Extrapolated tracks must be in the vicinity of the
$\epem$ interaction point, $i.e.$ within 1.5~cm in the plane
transverse to the beam axis and 2.5~cm along the beam axis. The
tracks used for the reconstruction of $\eta\ra\pip\pim\piz$ must
in addition have a transverse momentum $p_T$ larger than 50~$\mevc$.
When PID criterion is required on a track, the track polar angle
$\theta$ must be in the DIRC fiducial region
$25.78^{\circ}<\theta<146.10^{\circ}$. Photons are defined as
single bumps in the EMC crystals not matched with any track, and with a
shower lateral shape consistent with a photon. Because of high
background in the forward region of the EMC caused by the beam
asymmetry, the photons detected in the region
$\theta<21.19^{\circ}$ are rejected.

Intermediate resonances of the decays $\Bzb\ra\Dstze\hz$ are
reconstructed by combining tracks and$/$or photons for the
channels with the highest decay rate and detection efficiency.
When mesons are combined to build a resonance, masses are fixed to their nominal value~\cite{ref:PDG}.
For the $\omega$ and $\rho^0$ mesons, whose
natural width is not negligible in comparison with the experimental
resolution, their mass is not fixed to the nominal value.
The selections applied to each meson
$\piz$, $\eta$, $\omega$, $\etapr$, $\Dz$, and $\Dstarz$ are
optimized by maximizing the figure of merit
$S/\sqrt{S+B}$ where $S$ is the number of signal and $B$ is the
number of background events. The numbers $S$ and $B$ are computed
from simulations, the $\BF$'s used to evaluate $S$ are the world
average value given by PDG~\cite{ref:PDG}. Each resonance mass
distribution is fitted with a set of Gaussian functions or
a {\it modified Novosibirsk} function~\cite{ref:bukin}, which is composed of a Gaussian-like peaking
part with two tails at low and high values. Resonance candidates are then
required to have a mass within $\pm 2.5~\sigma$ around the fitted
mass central value, where $\sigma$ is the resolution of the mass
distribution obtained by the fit. For the resonances
$\Dz\ra\Km\pip\piz$ and $\Dstarz\ra\Dz\g$, the lower bound is
extended to $-3\sigma$ because of the photon energy losses in front and between the EMC crystals, which
makes the mass distribution asymmetric with a tail at low values.

\subsection{Selection of intermediate resonances}

\subsubsection{$\piz$ selection}\label{pizselection}

The $\piz$ mesons are reconstructed by combining two photons, each
photon energy $E(\gamma)$ must be larger than 85~$\mev$ for $\piz$
coming from $\Bz$ decay, and larger than 60~$\mev$ for $\piz$ coming
from $\eta$, $\omega$ or $\Dz$. Soft $\piz$'s coming from
$\Dstarz\ra\Dz\piz$ must satisfy $E(\gamma)>30~\mev$. The
reconstruction resolution for high momentum $\piz$ is limited by
the angle between the two daughter photons; for low momentum
$\piz$ the resolution is limited by the neutral hadron background
in EMC. The $\piz$ reconstructed mass resolution is about $6~\mevcc$ for
$\piz$ coming from $\eta$, $7~\mevcc$ for $\piz$ coming from
$\omega$ or $\Dz$, and $8~\mevcc$ for $\piz$ coming from $\Bz$ or
$\Dstarz$. The selection efficiency on signal ranges from 85 to
93~$\%$.

\subsubsection{$\eta$ selection}\label{etaselection}

The $\eta$ mesons are reconstructed in the $\gg$ and
$\pip\pim\piz$ decay modes. These modes account for 62~$\%$ of the total
decay rate~\cite{ref:PDG}, and may originate from
$\Bzb\ra\Dstze\eta$ or $\etapr\ra\pip\pim\eta$ decays.

The $\eta\ra\gg$ candidates are reconstructed by combining two
photons that satisfy $E(\gamma)>200~\mev$ for $\Bzb$ daughters and
$E(\gamma)>180~\mev$ for $\etapr$ daughters. As high momentum
$\piz$s may fake $\eta\ra\gg$ decays, a veto is applied against
$\piz$: for each $\eta\ra\gamma_1\gamma_2$ candidate, the photons
$\gamma_{1/2}$ are associated with photons of the rest of event
$\gamma_i$. If $E(\gamma_i)>200~\mev$ and the invariant mass of
the pair $\{\gamma_1\gamma_i\}$ or $\{\gamma_2\gamma_i\}$ is in
the $\piz$ mass window $115<m(\gamma_{1/2}\gamma_i)<150~\mevcc$,
then the $\eta$ candidate is rejected. The resolution of the
$\eta\ra\gg$ mass distribution is dominated by the EMC resolution
and is about $15~\mevcc$.

For $\eta$ candidates reconstructed in $\pip\pim\piz$, the $\piz$
is required to satisfy the conditions described in
Section~\ref{pizselection}. The mass resolution is about $3.5~\mevcc$,
which is smaller than for $\eta\ra\gg$ thanks to the higher
resolution of the tracking system. The selection efficiency on
signal is about 77~$\%$ for $\eta\ra\gg$ and 75~$\%$ for
$\eta\ra\pip\pim\piz$.

\subsubsection{$\omega$ selection}

The $\omega$ mesons are reconstructed in the $\pip\pim\piz$ decay
mode. These modes account for 89~$\%$ of the total decay rate. The $\piz$s
are required to satisfy the conditions described in
Section~\ref{pizselection} and the $\pipm$ must fulfill the
condition $p_T(\pipm)>200~\mevc$. The natural width of the
$\omega$ mass distribution $\Gamma\sim 8.49~\mev$~\cite{ref:PDG}
is comparable to the experimental resolution $\sigma\sim7~\mevcc$,
therefore the $\omega$ mass is not constrained to its nominal
value. We define a total width
$\sigma_{tot}=\sqrt{\sigma^2+\Gamma^2}\sim 11~\mevcc$ and require
the $\omega$ candidates to satisfy
$|m(\omega)-m(\omega)_{\textrm{mean}}|<2.5~\sigma_{tot}$. The
selection efficiency on signal is about 82~$\%$.

\subsubsection{$\rho^0$ selection}\label{rhoselection}

The $\rho^0$ mesons originate from $\etapr\ra\rho^0\g$ and are
reconstructed in the $\pip\pim$ decay mode. These modes account for
100~$\%$ of the total decay rate. Charged particle tracks must
satisfy $p_T(\pipm)>100~\mevc$. We define the helicity angle
$\theta_{\rho^0}$ as the angle between the pion momentum in the
$\rho^0$ rest frame and the $\rho^0$ momentum in the $\etapr$ rest
frame. Because the $\rho^0$ is a vector and pion is a pseudoscalar,
the angular distribution is proportional to
$\sin(\theta_{\rho^0})^2$ for pure signal and is flat for
background. The $\rho^0$ candidates with
$|\cos(\theta_{\rho^0})|>0.73$ are rejected. Due to the $\rho^0$
large natural width $\Gamma\sim149.4~\mev$~\cite{ref:PDG}, no
mass constraint is applied to the $\rho^0$. The mass of the
$\rho^0$ candidates must be within 160~$\mevcc$ around the nominal
mass value.

\subsubsection{$\etapr$ selection}

The $\etapr$ mesons are reconstructed in the
$\pip\pim\eta(\ra\gg)$ and $\rho^0\g$ decay modes. These modes account for
30~$\%$ of the total decay rate. For the reconstruction of
$\etapr\ra\pip\pim\eta(\ra\gg)$, the $\eta$ candidates must
satisfy the selections described in Section~\ref{etaselection}. The $\etapr$ mass resolution is about
$3~\mevcc$.

For $\etapr$ candidates reconstructed in $\rho^0\g$, the
$\rho^0$'s are required to satisfy the conditions described in
Section~\ref{rhoselection} and the photons must have an energy
larger than 200~$\mev$. As photons coming from $\piz$ decays may fake
signal, the veto against $\piz$ described in Section
\ref{etaselection} is applied. The $\etapr$ mass resolution is about
8~$\mevcc$, which is worse than $\etapr\ra\pip\pim\eta$ because of
the resolution on gamma reconstruction and the large $\rho$ mass
width.

The selection efficiency on signal is about 69~$\%$ for
$\etapr\ra\pip\pim\eta(\ra\gg)$ and 66~$\%$ for
$\etapr\ra\rho^0\g$.

\subsubsection{$\KS$ selection}\label{KSselection}

The $\KS$ mesons are reconstructed in the $\pim\pip$ decay modes. These modes account for 69~$\%$ of the total decay rate. The $\chi^2$
probability of the vertex fit of charged pions must be larger than
0.1~$\%$. We define the flight significance as the ratio
$L/\sigma_L$ where $L$ is the $\KS$ flight length in the plane
transverse to the beam axis and $\sigma_L$ is the uncertainty on
$L$ determined from the vertex fit constraint. The combinatorial
background is rejected by requiring a flight significance larger than 5.
The reconstructed $\KS$ mass resolution is about 2~$\mevcc$. The
selection efficiency on signal is about 86~$\%$.

\subsubsection{$\Dz$ selection}\label{Dzselection}

The $\Dz$ mesons are reconstructed in $\Km\pip$, $\Km\pip\piz$,
$\Km\pip\pim\pip$, and $\KS\pip\pim$ decay modes. These modes account for about
28~$\%$ of the total decay rate. All $\Dz$ candidates must satisfy
$p^*(\Dz)>1.1~\gevc$. That requirement is loose enough that background populate the sidebands of the signal region. The $\pipm$'s coming from the
$\Dz$ candidate must fulfill $p_T(\pipm)>400~\mevc$ for
$\Dz\ra\Km\pip$, $p_T(\pipm)>100~\mevc$ for
$\Dz\ra\Km\pip\pim\pip$ and $p_T(\pipm)>120~\mevc$ for
$\Dz\ra\KS\pim\pip$. The $\chi^2$ probability of the vertex fit of
charged pions must be larger than 0.1~$\%$ for $\Dz\ra\Km\pip$ and larger than
0.5~$\%$ for the other modes. The kaon candidates must satisfy
tight kaon criteria for $\Km\pip$ and $\Km\pip\piz$, and a loose
kaon criteria for $\Km\pip\pim\pip$. For $\KS\pip\pim$, the $\KS$
candidates must satisfy the selection criteria described in
Section~\ref{KSselection}.

The decay $\Dz$ to $\Km\pip\piz$ proceeds mainly through the
resonances $\Kstar$ ($\Kstarz\ra\Km\pip$ or $\Kstarm\ra\Km\piz$)
and $\rho^+(\pip\piz)$. Combinatorial background is rejected by
using the parametrization of the $\Km\pip\piz$ Dalitz distribution
studied by the Fermilab E691 experiment~\cite{ref:E691}. The $\Dz$
candidates that are not in the resonance regions of the Dalitz
distribution are rejected. The $\piz$ must satisfy the selections
described in Section~\ref{pizselection}.

The reconstructed $\Dz$ mass resolution is about 4, 5, 6, and
11~$\mevcc$ for $\Km\pip\pim\pip$, $\KS\pip\pim$, $\Km\pip$, and
$\Km\pip\piz$ modes, respectively. To account for the asymmetry in
the mass distribution of $\Dz\ra\Km\pip\piz$, we require that the $\Dz$ candidates satisfy
$-3~\sigma<(m(\Dz)-m(\Dz)_{\textrm{mean}})<2.5~\sigma$.

The selection efficiency on signal is about 71~$\%$ for $\Km\pip$,
60~$\%$ for $\Km\pip\pim\pip$, 71~$\%$ for $\KS\pip\pim$, and
44~$\%$ for $\Km\pip\piz$.

\subsubsection{$\Dstarz$ selection}

The $\Dstarz$ mesons are reconstructed in the $\Dz\piz$ and
$\Dz\g$ decay modes. The $\piz$ and $\Dz$ candidates are required
to satisfy the selections described in Section~\ref{pizselection} and \ref{Dzselection} respectively. The
photons from $\Dstarz\ra\Dz\g$ must fulfill $E(\g)>200~\mev$ and
pass the veto against $\piz$ described in Section~\ref{etaselection}.

The resolution of the mass difference $\Delta m\equiv
m(\Dstarz)-m(\Dz)$ is about 2~$\mevcc$ for $\Dz\piz$ and
7~$\mevcc$ for $\Dz\g$. The $\Dstarz\ra\Dz\piz$ candidates must
satisfy $|\Delta m-\Delta m_{\textrm{mean}}|<2.5~\sigma$. Because
of the asymmetry of the $\Delta m$ distribution for $\Dz\g$, the
$\Dstarz$ candidates must satisfy $-3~\sigma<(\Delta m-\Delta
m_{\textrm{mean}})<2.5~\sigma$.

The selection efficiency on signal is about 49~$\%$ for $\Dz\piz$
and 39~$\%$ for $\Dz\g$.

\subsection{Selection of $B$ candidates}

The $B$ candidates are reconstructed by combining a $\Dstze$ with
an $\hz$, with the $\Dstarz$ and $\hz$ masses constrained to their
nominal value except when $\hz$ is the $\omega$. One needs to
discriminate real $B$'s from fake ones created from combinatorial
or crossfeed background.

\subsubsection{$B$ kinematic variables}

Two kinematic variables are used in $\babar$ to select $B$
candidates: the energy-substituted mass $\mes$ and the energy
difference $\DeltaE$. These two variables use the constraints from
the precise knowledge of the beams' energies and from energy
conservation in the two-body decay $\Upsilon(4S)\ra\BB$. The
quantity $\mes$ is the invariant mass of the $B$ candidate where
the $B$ energy is set to the beam energy in the CM frame:
\begin{equation}
\mes =
\sqrt{\left(\frac{\sqrt{s}}{2}\right)^2-|\overrightarrow{p^*_B}|^2},
\end{equation}
and $\DeltaE$ is the energy difference between the reconstructed
$B$ energy and the beam energy in the CM frame:

\begin{equation}
\DeltaE = E^*_B - \sqrt{s}/2,
\end{equation}
where $\sqrt{s}$ is the $\epem$ CM energy and
$(E^*_B,\overrightarrow{p_B})$ is the $B$ quadrivector in the CM
frame. For $B$ signal events, the $\mes$ distribution peaks at the
$B$ mass with a resolution of about 3~$\mevcc$ dominated by the
beam energy spread, whereas $\DeltaE$ peaks near zero with a
resolution of $15-50~\mev$ depending on the number of photons in
the final state.

\subsubsection{Rejection of $\epem\ra\qqbar$ background}

The continuum background $\epem\ra\qqbar$ ($q=u$, $d$, $s$ or mainly
$c$) creates real high momentum mesons $\Dstze$, $\piz$,
$\eta^{(')}$, $\omega$ that can fake the signal mesons originating
from the two body decays $\Bzb\ra\Dstze\hz$. That background is
thus not rejected by the selections on intermediate resonances.
Since the $B$ mesons are produced almost at rest in the $\Upsilon(4S)$
frame, so the $\Upsilon(4S)\ra\BB$ events shape is spherical. By
comparison, the $\qqbar$ events have a back-to-back jet-like shape
given that the quark masses ($q\ne b$) are small compared to the
CM energy. The $\qqbar$ background can hence be discriminated by
the event shape described by the following variables:

\begin{itemize}

\item The thrust angle $\theta_T$ defined as the angle between the
thrust axis of the $B$ candidate and the thrust axis of the rest
of event. The thrust axis $\widehat{T}$ is defined as the axis
that maximizes the quantity $T$:
\begin{equation}
T = \frac{\sum_i|\overrightarrow{p_i}\cdot\widehat{T}|}{\sum_i
|\overrightarrow{p_i}|}.
\end{equation}
The distribution of $|\cos(\theta_T)|$ is flat for signal and
peaks at $1$ for continuum background.

\item Legendre monomials $L_0$ and $L_2$ defined as:
\begin{equation}
L_0 = \sum_i p^*_i\ ;\ L_2 = \sum_i p^*_i|\cos(\theta^*_i)|^2,
\end{equation}

with $p_i^*$ the momentum of the
particle $i$  that does not come from a $B$ candidate, and
$\theta_i^*$ is the angle between $p_i^*$ and the thrust axis of
the $B$ candidate.

\item The polar angle $\theta^*_B$ between the $B$ momentum in the
$\Upsilon(4S)$ frame and the beam axis. The $\Upsilon(4S)$ being
vector ($J=1$) and the $B$ mesons being pseudoscalar ($J=0$), the
angular distribution is proportional to $\sin^2(\theta^*_B)$ for
signal and roughly flat for background.

\end{itemize}

These four variables are combined in a Fisher discriminant built
with the $T$MVA~\cite{ref:TMVA} toolkit package. The Fisher
$\mathcal{F}_{shape}$ is trained with signal MC events and
off-peak data events; in order to maximize the number of off-peak
events all the $\Bzb\ra\Dstze\hz$ modes are combined. The training
and test are performed with $2\times 20000$ signal events and
$2\times 20000$ off-peak events. The obtained Fisher formula is:
\begin{eqnarray}
\mathcal{F}_{shape} =
2.36-1.18\cdot|\cos(\theta_T)|+\nonumber\\
0.20\cdot L_0 - 1.01\cdot L_2 - 0.80\cdot |\cos(\theta_B^*)|.
\end{eqnarray}
The $\qqbar$ background is rejected by applying a selection on
$\mathcal{F}_{shape}$. The selection is optimized for each $\Bzb$
signal mode by maximizing the statistical significance with signal
MC against generic MC $\epem\ra\qqbar$, $q\ne b$. That requirement
retains between 36~$\%$ and 98~$\%$ of $B$ signal, while rejecting
between 23$\%$ and 97~$\%$ of the $\qqbar$ background respectively.

\subsubsection{Rejection of $\Upsilon(4S)\ra\BB$
background}\label{bruit_BB}

The $\omega$ mesons in $\Bzb\ra\Dz\omega$ are polarized. We define the
{\it normal angle}
$\theta_N$~\cite{ref:Babar2004,ref:BermanJacob} as the angle
between the normal to the $\omega$ decay plane and the direction
of $\Bzb$ in the $\omega$ rest frame. That definition is the
equivalent of the two-body helicity angle for the three-body
decay.  To describe the 3-body decay distribution of
$\omega\ra\pip\pim\piz$, we define the {\it Dalitz angle}
$\theta_D$~\cite{ref:Babar2004} as the angle between the $\piz$
momentum in the $\omega$ frame and the $\pip$ momentum in the
$\{\pip\pim\}$ frame.

Given the angular momenta of the $\Bzb$ and $\Dz$ mesons
($J=0$), and $\omega$ meson ($J=1$), the signal distribution is
proportional to $\cos^2(\theta_N)$ and $\sin^2(\theta_D)$ while
the combinatorial background distribution is roughly flat. These
two angles are combined in a Fisher discriminant
$\mathcal{F}_{hel}$ built from signal MC events and generic
$\qqbar$ and $\BB$ MC events:
\begin{equation}
\mathcal{F}_{hel}=-1.41-1.01\cdot|\cos(\theta_D)|+3.03\cdot|\cos(\theta_N)|.
\end{equation}
We require $\Bzb\ra\Dz\omega$ candidates to satisfy
$\mathcal{F}_{hel}>-0.1$, to obtain an efficiency on signal
(background) of 85~$\%$ (39~$\%$).

We also use the angular distribution of $\Dstarz\ra\Dz\piz$ to
reject combinatorial background. We define the helicity angle
$\theta_{D^*}$ as the angle between the $\Dz$ momentum in the
$\Dstarz$ frame and the $\Dstarz$ momentum in the $\Bzb$ frame.
The angular distribution is proportional to $\cos(\theta_{D^*})^2$
for signal and roughly flat for combinatorial background.
Selection on $|\cos(\theta_{D^*})|$ significantly improves the
statistical significance for the $\Bzb\ra\Dstarz\piz$ mode only.
The $\Dstarz$ candidates coming from the decay
$\Bzb\ra\Dstarz\piz$ are then required to satisfy
$|\cos(\theta_{D^*})|>0.4$ for an efficiency on signal of 90~$\%$
and on background of 62~$\%$

An important $\BB$ background contribution in the reconstruction
of $\Bzb\ra\Dstze\piz$ comes from the color-allowed decay
$\Bm\ra\Dstze\rho^-$. If the charged pion from the decay
$\rho^-\ra\pim\piz$ is not used in the construction of the $B$
candidate, $\Dstze\rho^-$ events mimic the $\Dstze\piz$
signal. Moreover, the $\BF$ of $\Bm\ra\Dstze\rho^-$ is almost 50
times larger than the signal $\BF$, and is known with a relative
total uncertainty of 13~$\%$ for $\Dz\rho^-$ and
17~$\%$ for $\Dstarz\rho^-$~\cite{ref:PDG}. A veto is applied
against $\Bm\ra\Dstze\rho^-$ events. For each $\Dstze\piz$
candidate, a $\Bm$ candidate is reconstructed in $\Dstze\rho^-$.
If the $\Bm$ candidate mass is in the signal region
$\mes(\Bm)>5.27~\gevcc$, $|\DeltaE(\Bm)|<100~\mev$ and
$|m(\rho^-)-m(\rho^-)_{\textrm{PDG}}|<250~\mevcc$, and if the
$\piz$ and $\Dz$ candidates are the same used to build the $\Bzb$
candidate, then the $\Bzb$ is rejected. For the reconstruction of
$\Dz\piz$, the veto retains about 90~$\%$ of signal and rejects
about $67~\%$ of $\Dz\rho^-$ and $44~\%$ of $\Dstarz\rho^-$. For
the reconstruction of $\Dstarz\piz$, the veto efficiency on signal
is 80~$\%$ while rejecting about 56 and 66~$\%$ on $\Dz\rho^-$ and $\Dstarz\rho^-$ respectively.

\subsubsection{Choice of one $B$ candidate}

The average number of $B$ candidate per event after all selections
ranges between 1 and 1.6 depending on the subdecays. The highest multiplicities correspond to
modes with the largest number of neutral particles in the final state. We keep one
$B$ candidate per mode per event. The chosen $B$ is the one with
the smaller value of:

\begin{eqnarray}
\chi^2_m
&=&\left(\frac{m(\Dz)-m(\Dz)_{\textrm{mean}}}{\sigma_{m(\Dz)}}\right)^2\nonumber\\
&+&\left(\frac{m(\hz)-m(\hz)_{\textrm{mean}}}{\sigma_{m(\hz)}}\right)^2,
\end{eqnarray}
for $\Dz\hz$ modes and

\begin{eqnarray}
\chi^2_m &=&
\left(\frac{m(\Dz)-m(\Dz)_{\textrm{mean}}}{\sigma_{m(\Dz)}}\right)^2
\nonumber\\
&+&\left(\frac{m(\hz)-m(\hz)_{\textrm{mean}}}{\sigma_{m(\hz)}}\right)^2\nonumber\\
&+&\left(\frac{\Delta m-\Delta m_{\textrm{mean}}}{\sigma_{\Delta
m}}\right)^2,
\end{eqnarray}
for the $\Dstarz\hz$ modes. The quantities $\sigma_{m_{\Dz}}$ and
$\sigma_{m_{\hz}}$ (with respect to $m(\Dz)_{\textrm{mean}}$ and
$m(\hz)_{\textrm{mean}}$) are the resolutions (with respect to means)
of the reconstructed mass distributions. The quantities $\Delta
m_{\textrm{mean}}$ and $\sigma_{\Delta m}$ are respectively the
mean and resolution of the $\Delta m$ distribution. These quantities
are obtained from fits of the mass distribution of true simulated
candidates selected from signal MC simulations. An associated systematic error is calculated later because of the use of MC to calculate $\Delta
m_{\textrm{mean}}$ and $\sigma_{\Delta m}$. The efficiency on signal of the choice of one $B$ per event ranges from 76 to
100~$\%$. The lower efficiencies correspond to the modes with high
neutral multiplicity.

\subsubsection{MC efficiency corrections}\label{sec:MCcorrection}

The branching fraction is computed as:

\begin{equation}\label{eq:BF}
\BF=\frac{S}{n_{B\bar{B}}\cdot\epsilon\cdot\BF_{sec}},
\end{equation}
where $\BF_{sec}$ is the product of the $\BF$'s of the
intermediate resonances reconstructed decays, which are taken from
the PDG~\cite{ref:PDG}, $n_{B\bar{B}}$ is the number of $\BB$
pairs in data and $S$ is the number of signal events remaining
after all selections. The quantity $\epsilon$ is the signal
efficiency of reconstruction and selections, computed from the
exclusive MC simulations, and the quantity
$\epsilon\cdot\BF_{sec}$ is the total acceptance. When computing
the $\BF$ of the sum of $\Dstze$ subdecays, the corresponding
total acceptances are summed. The total acceptance is corrected by about 1~$\%$ (10~$\%$) from the crossfeed between $\Dz$
($\Dstarz$) decay modes. The high correction for the sum of
$\Dstarz$ decays is mainly due to the large crossfeed of
$\Dstarz\ra\Dz\piz$ into $\Dstarz\ra\Dz\g$. That correction was obtained by computing with exclusive MC's the proportion of intern crossfeed by respect to real signal.

The selection efficiency from simulation is slightly different
from the efficiency in data. The MC efficiency for the
reconstruction of $\piz/\g$ is adjusted from the study of $\tau$
decays in the channels $\tau\ra\rho(\pi\piz)\nu_{\tau}$ and
$\tau\ra\pi\nu_{\tau}$~\cite{ref:neutralEff}. The correction on track efficiency is
computed from studies of track mis-reconstruction in the decays
$\tau\ra(\pip\pim)\mathbf{\pim}\nu_{\tau}$ and
$\tau\ra\rho^0(\pip\pim)\mathbf{\pim}\nu_{\tau}$~\cite{ref:trackingEff}. The simulated efficiency
of charged particle identification is compared to the efficiency
computed in data with pure samples of kaons from the decays
$D^{*+}\ra\Dz\pip$ and $\Dz\ra\Km\pip$. The efficiency on $\KS$
candidates is modified using a high statistics and a high purity data
sample of $\KS$ mainly arising from the process $\epem\ra\qqbar$.

The efficiency corrections for the selections applied to $\Dstze$
candidates and on the Fisher discriminant for the $\qqbar$
rejection are obtained from the study of the control sample
$\Bm\ra\Dstze\pim$. That control sample was chosen for its high
statistics and purity, and for its similarity with
$\Bzb\ra\Dstze\hz$. The correction is computed from the double
ratio eff(data)$/$eff(MC), where the efficiencies are computed
from $\mes$ fits of $\Bm\ra\Dstze\pim$ events in data and MC,
before and after the corresponding selections. The obtained
results were checked with the color-allowed control sample
$\Bm\ra\Dstze\rho^-$ which has a slightly different kinematic due to the mass of the $\rho^-$ and validates this correction for modes as $\Dstze\etapr$.

\subsection{Probability density functions (pdf) and fit procedure}

The signal yield $S$ is extracted by an extended unbinned maximum
likelihood (ML) fit of the $\DeltaE$ distribution in the range
$-0.280<\DeltaE<0.280~\gev$ for $\mes>5.27~\gevcc$. Fitting
$\DeltaE$ allows us to model and to fit the complex crossfeed structure without
relying on simulation completely.

Due to the energy loss by photons in the detector material before the
EMC, the $\DeltaE$ distribution for signal is modelled by a
modified Novosibirsk function~\cite{ref:bukin}. A Gaussian is added to
the modes with a large $\DeltaE$ resolution to describe the
mis-reconstructed events. The signal shape parameters are
estimated from a ML fit of simulated signal events in exclusive
decay modes.

The dominant crossfeed contribution for $\Bzb\ra\Dz\hz$ comes from
the channel $\Bzb\ra\Dstarz\hz$ when the $\piz/\g$ from the
$\Dstarz$ decay is not reconstructed. Such crossfeed events are
shifted in $\DeltaE$ by about $-135~\mev$ with a tail from
$\Dstarz(\ra\Dz\g)\hz$ in the signal region.

\begin{table}[htb]
\caption{\label{tab:crossfeed}Main crossfeed between signal
modes.}
\begin{center}
\begin{tabular}{lc}
\hline\hline
$B$ mode       &  Crossfeed mode  \\
\hline
$\Dz\hz$ & $\Dstarz\hz$ \\
$\Dstarz(\Dz\piz)\hz$ &  $\Dstarz(\Dz\g)\hz$, $\Dz\hz$\\
$\Dstarz(\Dz\g)\hz$ &  $\Dstarz(\Dz\piz)\hz$, $\Dz\hz$\\
$\Dstarz\hz$ &  $\Dz\hz$\\
\hline\hline
\end{tabular}
\end{center}
\end{table}
The main crossfeed contributions from the other reconstructed
$\Bz$ modes are shown in Table~\ref{tab:crossfeed}. For each
signal mode, different crossfeeds are summed and their
contribution is estimated with a histogram-based pdf built from
the signal MC. The major crossfeed in the reconstruction of
$\Bzb\ra\Dstze\piz$ comes from the decays $\Bm\ra\Dstze\rho^-$
(see Section~\ref{bruit_BB}). Their contribution is modelled by a
separate histogram-based pdf built from the exclusive MC
simulations.

The combinatorial background from $\BB$ and $\qqbar$ are summed
and modelled by a 2$^{nd}$ order polynomial. In the cases of
$\Bzb\ra\Dstze\omega(\eta)$ modes, an additional crossfeed
contribution comes respectively from the decays $\Bm\ra\Dstze\pim\omega(\eta)$.
In the reconstruction of $\Bzb\ra\Dz\piz$ where all $\Dz$ subdecays are summed, an additional crossfeed contribution comes
mainly from the decays $\Bm\ra\Dz\rho^-$, $\Bzb\ra\Dz\piz\piz$,
$\Bzb\ra\bar{K}^{*0}\piz\piz$ and $\Bzb\ra\Dz\piz$ where $\Dz$
decays to a \CP eigenstate or through a semileptonic channel.
These peaking backgrounds are significant when summing the $\Dz$
submodes. In these cases, the combinatorial background is
modelled by a first-order polynomial plus a Gaussian to take into
account additional crossfeeds. For $\Bzb\ra\Dstze\omega(\eta)$
modes, the fraction of that Gaussian is left floating in the
fit, since the $\BF$'s of $\Bm\ra\Dstze\pim\omega(\eta)$ are
not precisely known.

The shape parameters of the combinatorial background pdf are
obtained from ML fits on the $\BB$ and continuum MC, where all
signal and crossfeed events have been removed.

The pdf normalization for the signal, for the combinatorial background
and for the crossfeed $\Dstze\rho^-$ are allowed to float in the fit.
The mean of the signal pdf is left floating for the sum of
$\Dstze$ subdecays. For each $\Dz$ submodes the signal mean is
fixed to the value obtained with the fit on the sum of $\Dz$
submodes.

A given mode $\Bzb\ra\Dstarz\hz$ can be signal and
crossfeed to other modes at the same time. In order to use the
$\BF$ computed in this analysis, the yield extraction is performed
through an iterative fit on successively $\Dstarz\hz$ and
$\Dz\hz$. The normalization of crossfeed contribution from
$\Dstze\hz$ is then fixed to the $\BF$ measured in the previous
fit iteration. That iterative method converges quickly to a stable
value of $\BF$, with variation below 10~$\%$ of statistical uncertainty, in less than 5 iterations.

We check the absence of bias in our fit by studying embedded toy
MC's. The extraction procedure is applied to toy samples where
background events are generated from the fitted pdf's. The toy
signal events are taken from the corresponding exclusive MC with a
yield corresponding to the MC-generated value of the branching
fraction $\BF_{gen}$. No significant bias are found.

\subsection{Signal yield extraction}

\begin{figure*}[htb]
\begin{center}
\includegraphics[width=0.300\linewidth]{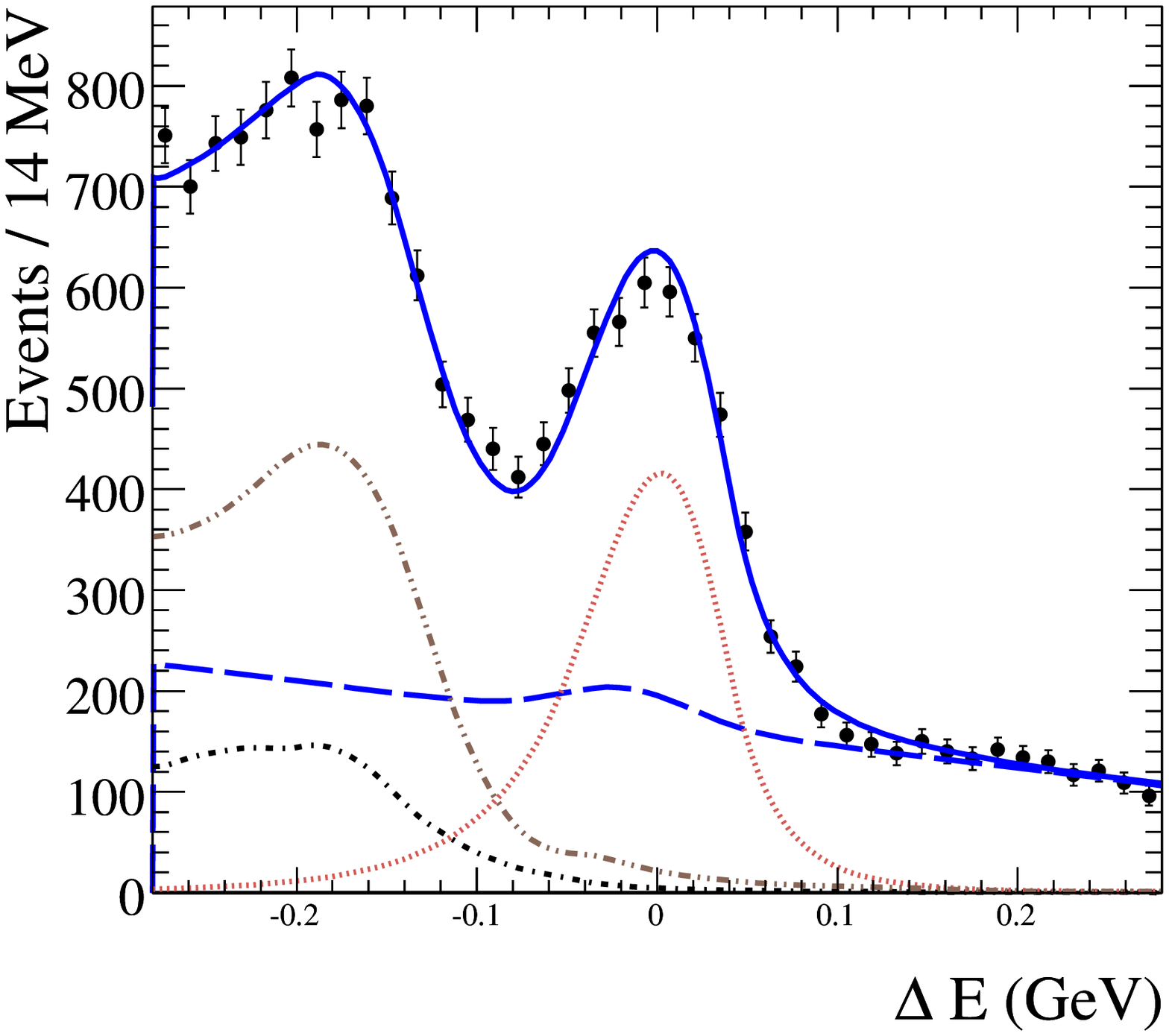}
\includegraphics[width=0.300\linewidth]{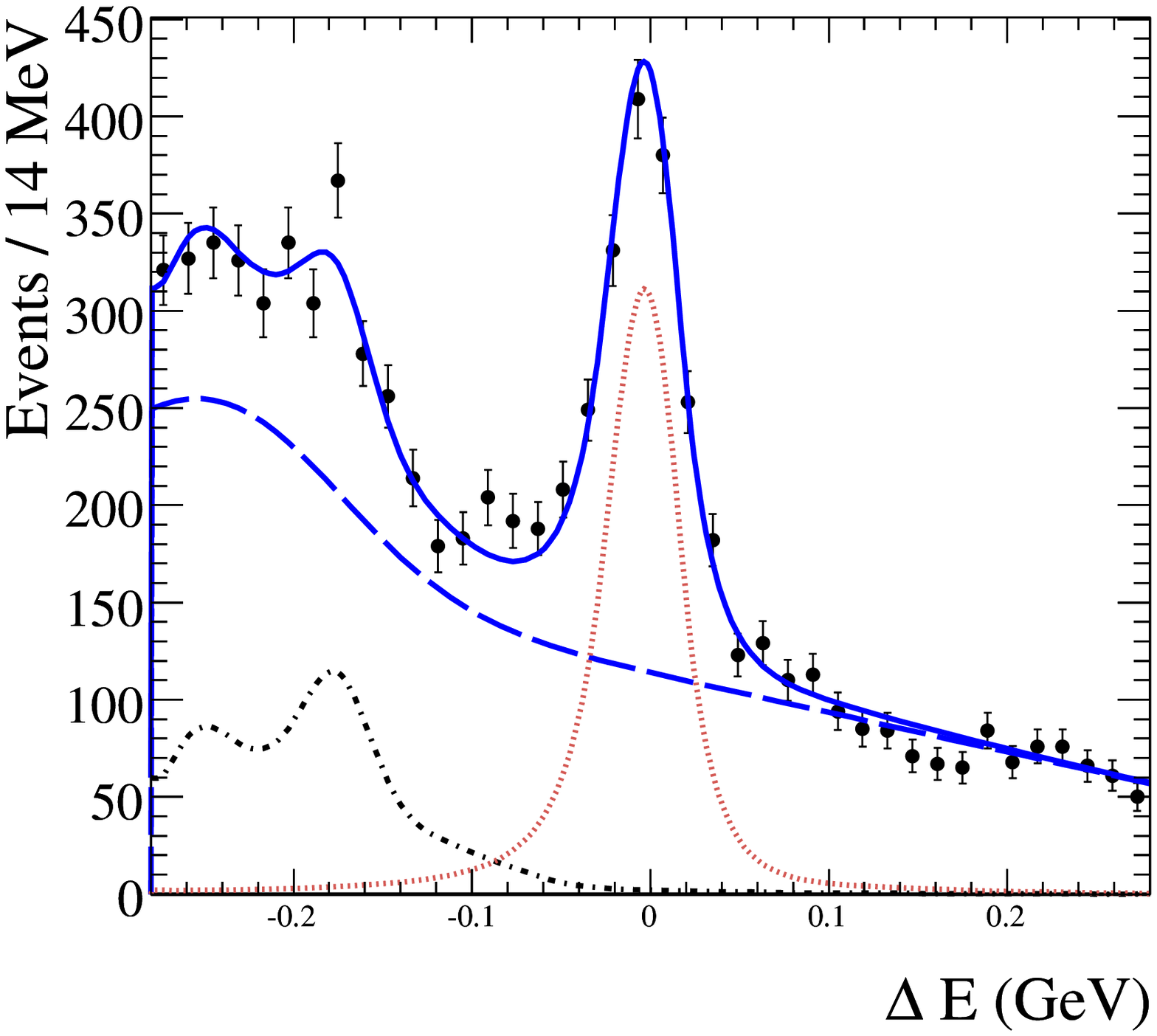}
\includegraphics[width=0.300\linewidth]{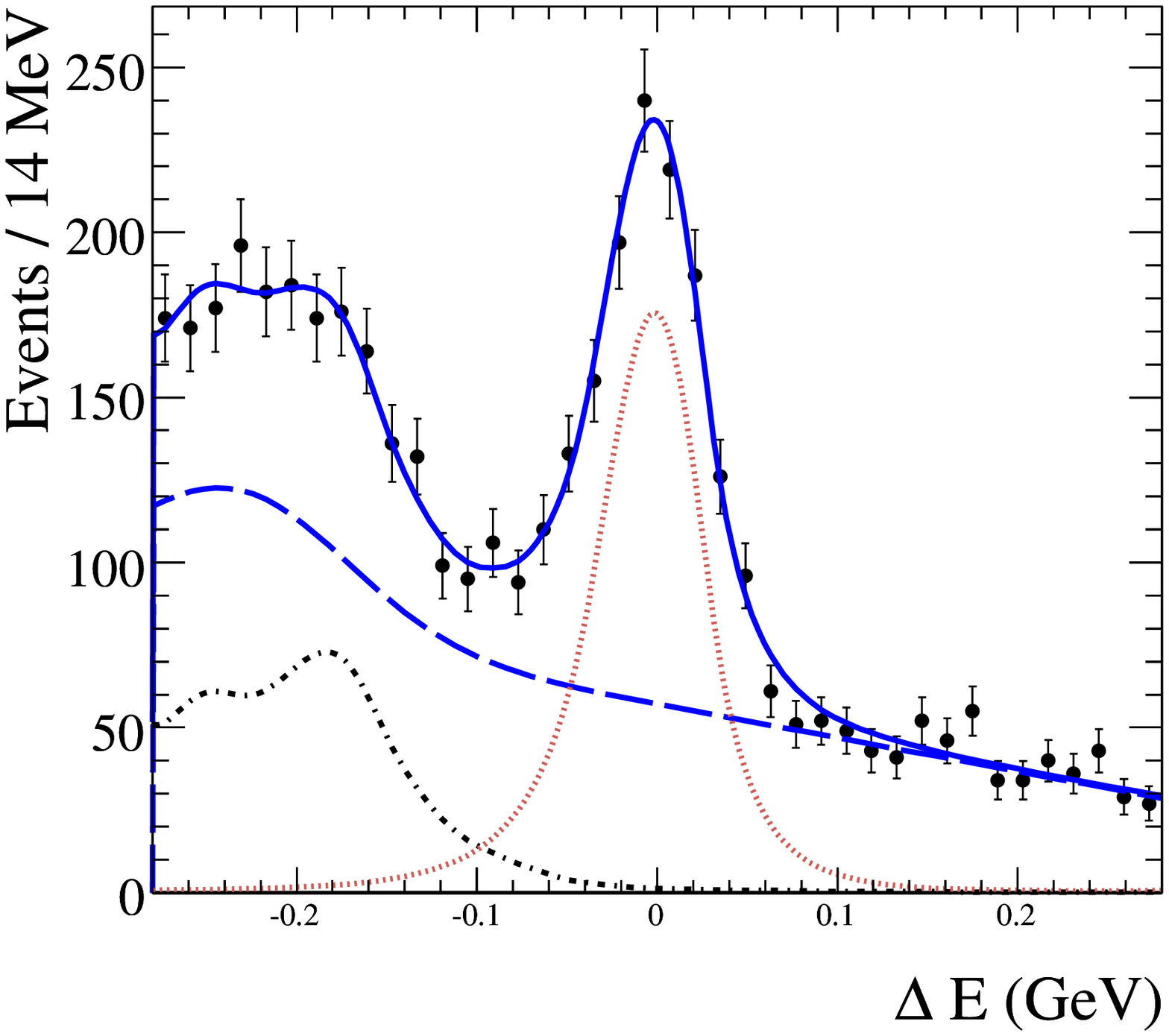}\\
\includegraphics[width=0.300\linewidth]{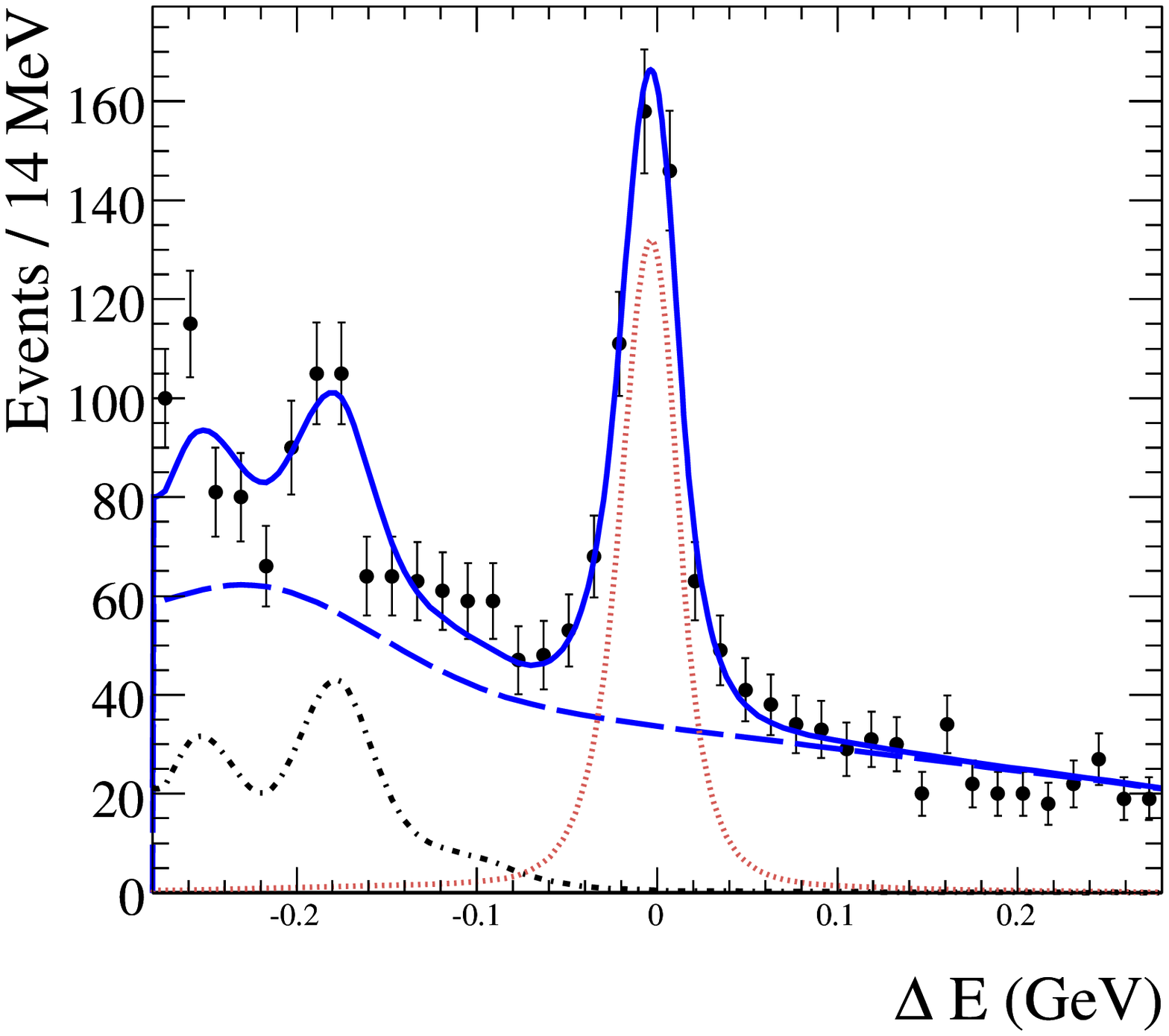}
\includegraphics[width=0.300\linewidth]{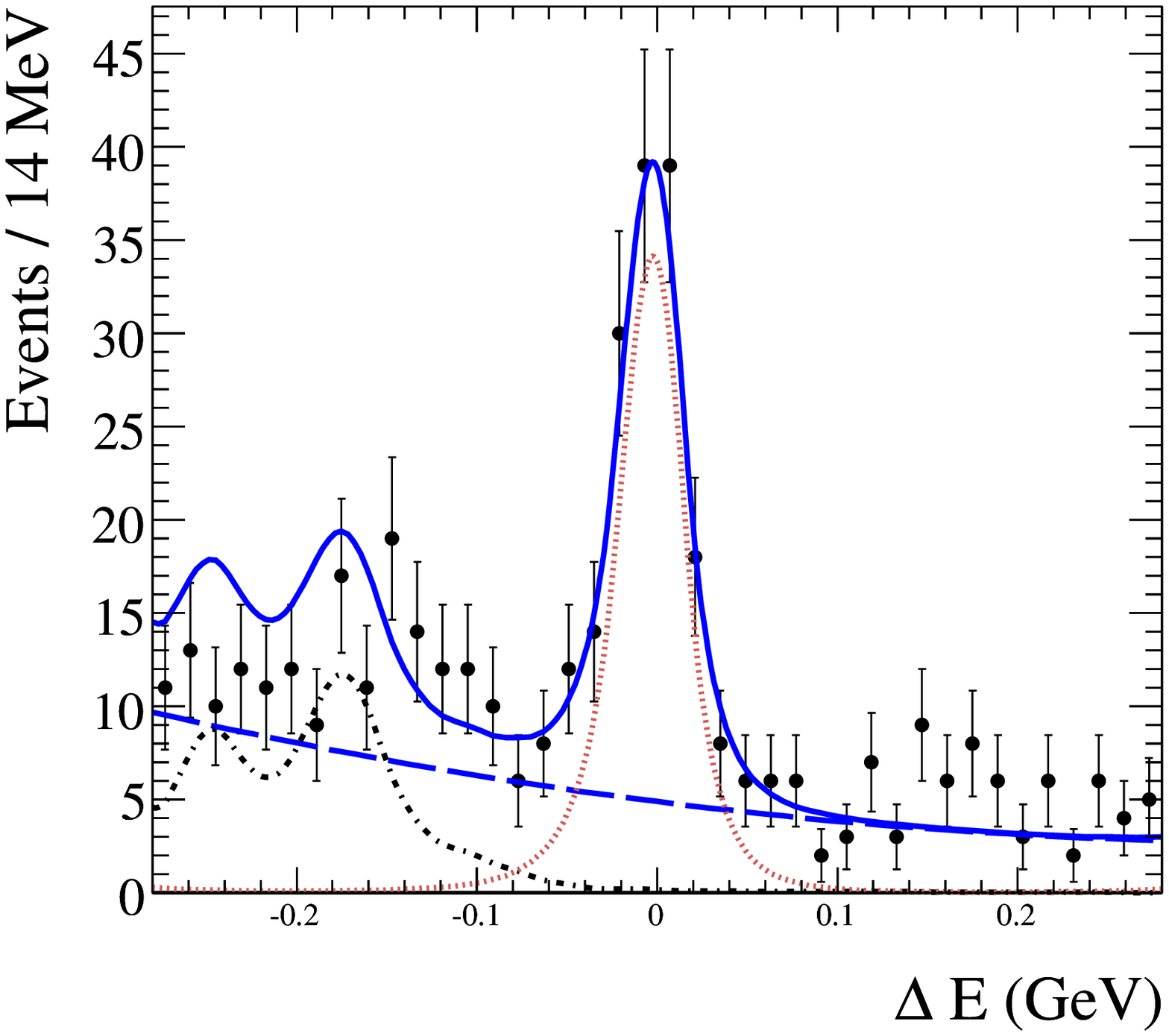}
\includegraphics[width=0.300\linewidth]{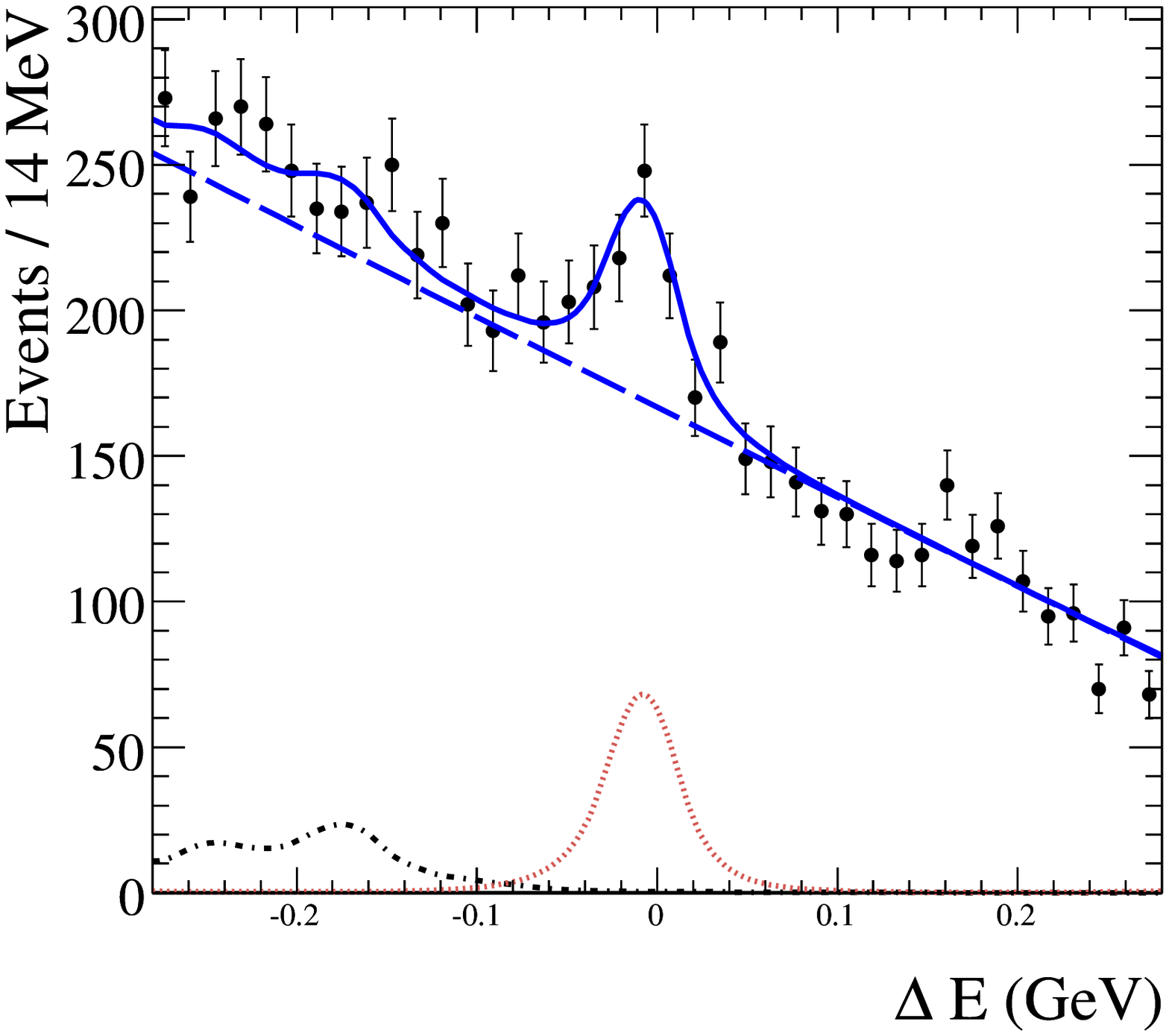}
\caption{\label{fig:fitData_1} Fit of $\DeltaE$ distributions in
data for modes $\Bzb\ra\Dz\piz$ (top left), $\Bzb\ra\Dz\omega$
(top middle), $\Bzb\ra\Dz\eta(\gg)$ (top right),
 $\Bzb\ra\Dz\eta(\pi\pi\piz)$ (bottom left), $\Bzb\ra\Dz\etapr(\pi\pi\eta)$
(bottom middle) and $\Bzb\ra\Dz\etapr(\rho^0\g)$ (bottom right).
The dots with error bars are data, the solid curve is the fitted
total pdf, the dotted curve is the signal pdf, the dotted-dashed
curve is the crossfeed pdf, the double dotted-dashed curve is the
$\Bm\ra\Dstze\rho^-$ pdf and the long dashed curve is the
combinatorial background pdf.}
\end{center}
\end{figure*}

\begin{figure*}[htb]
\begin{center}
\includegraphics[width=0.300\linewidth]{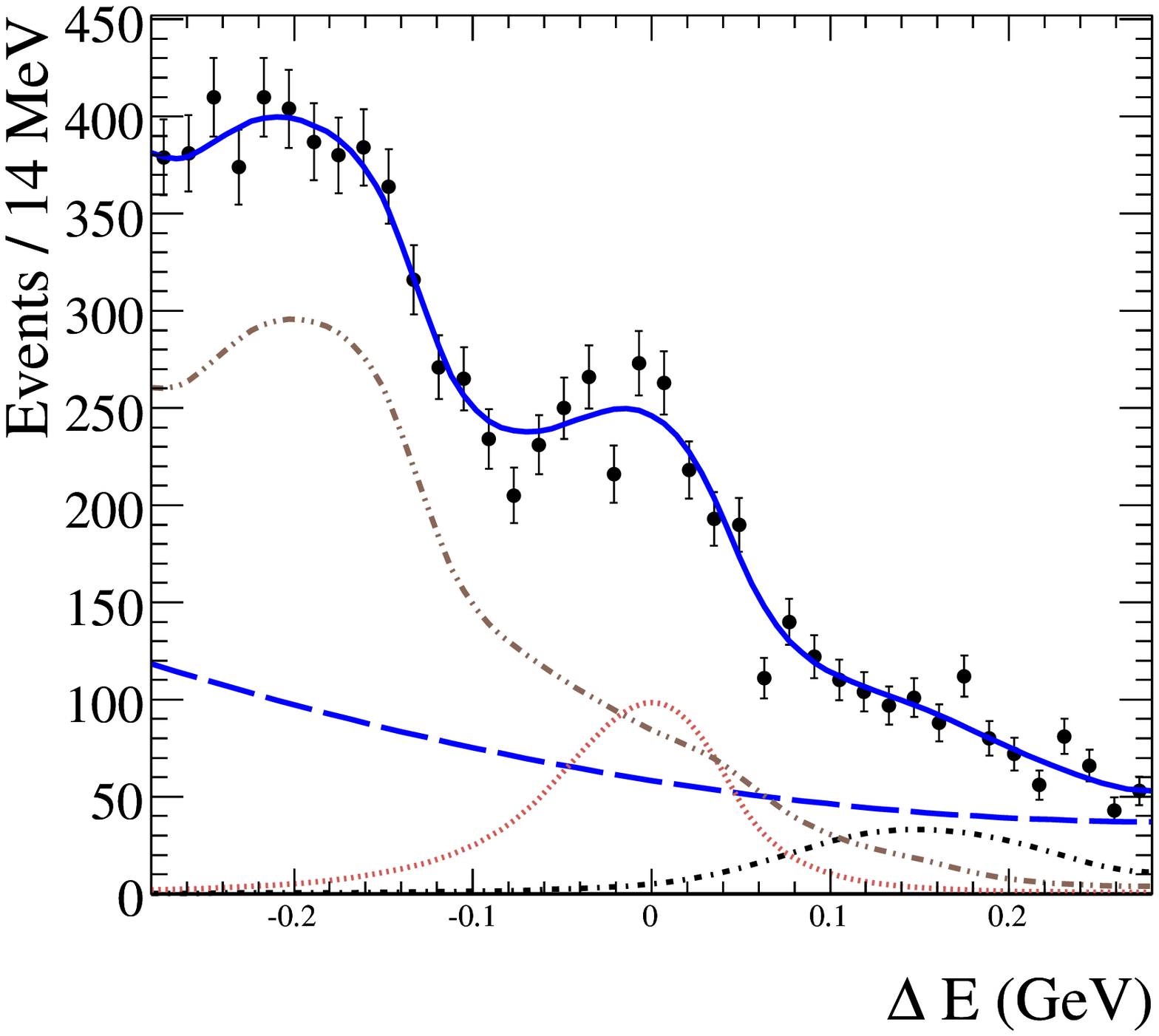}
\includegraphics[width=0.300\linewidth]{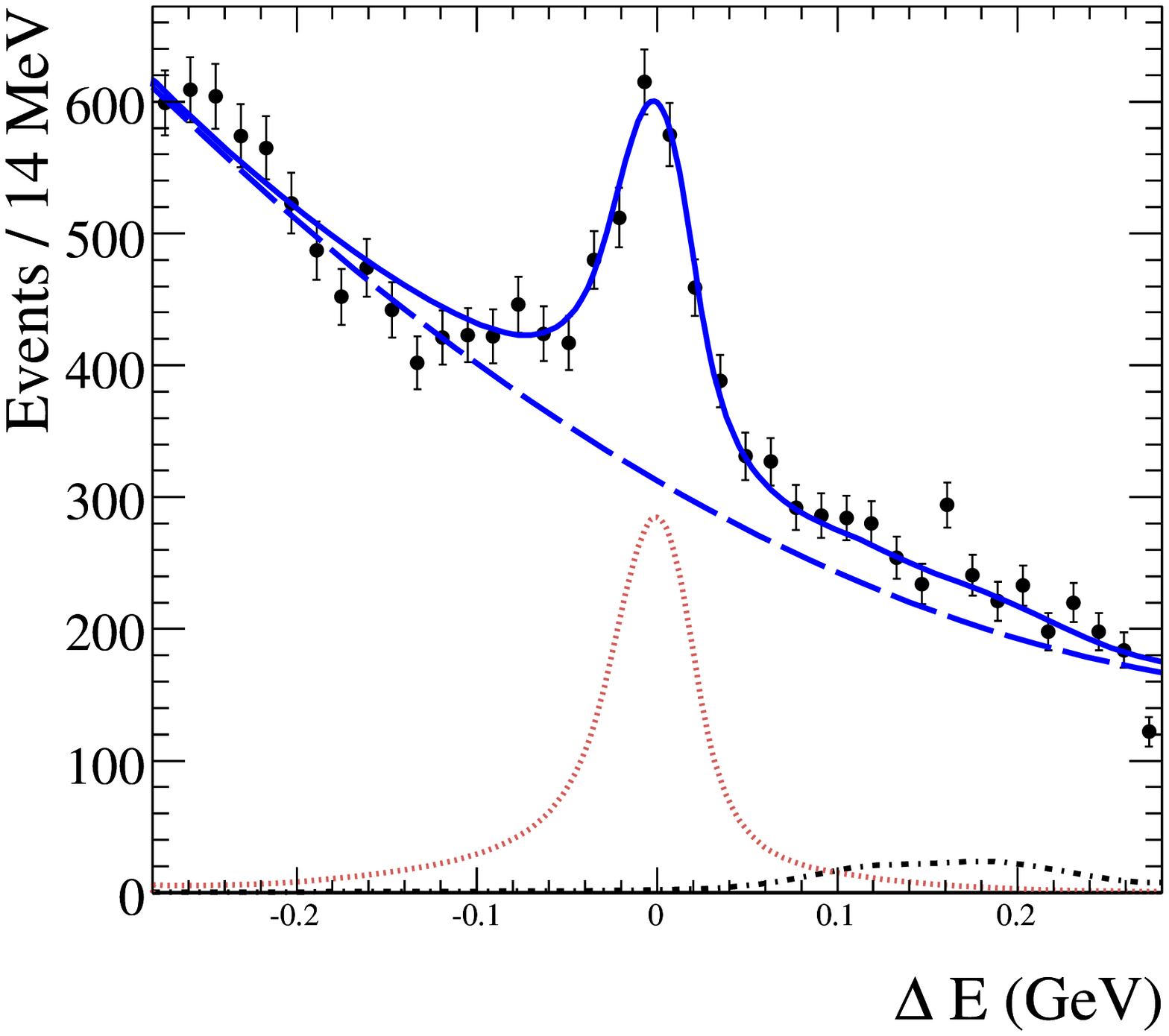}
\includegraphics[width=0.300\linewidth]{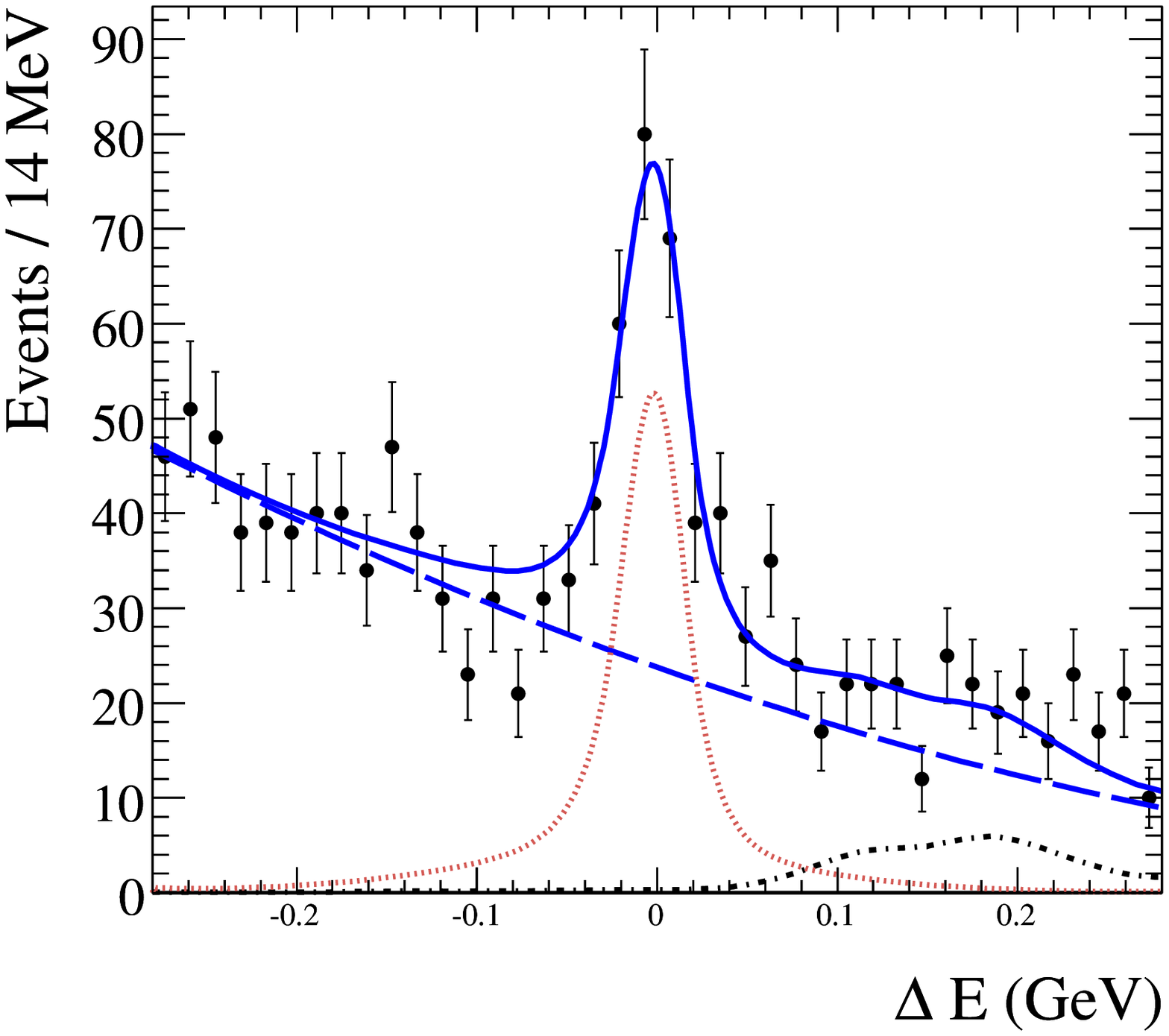}\\
\includegraphics[width=0.300\linewidth]{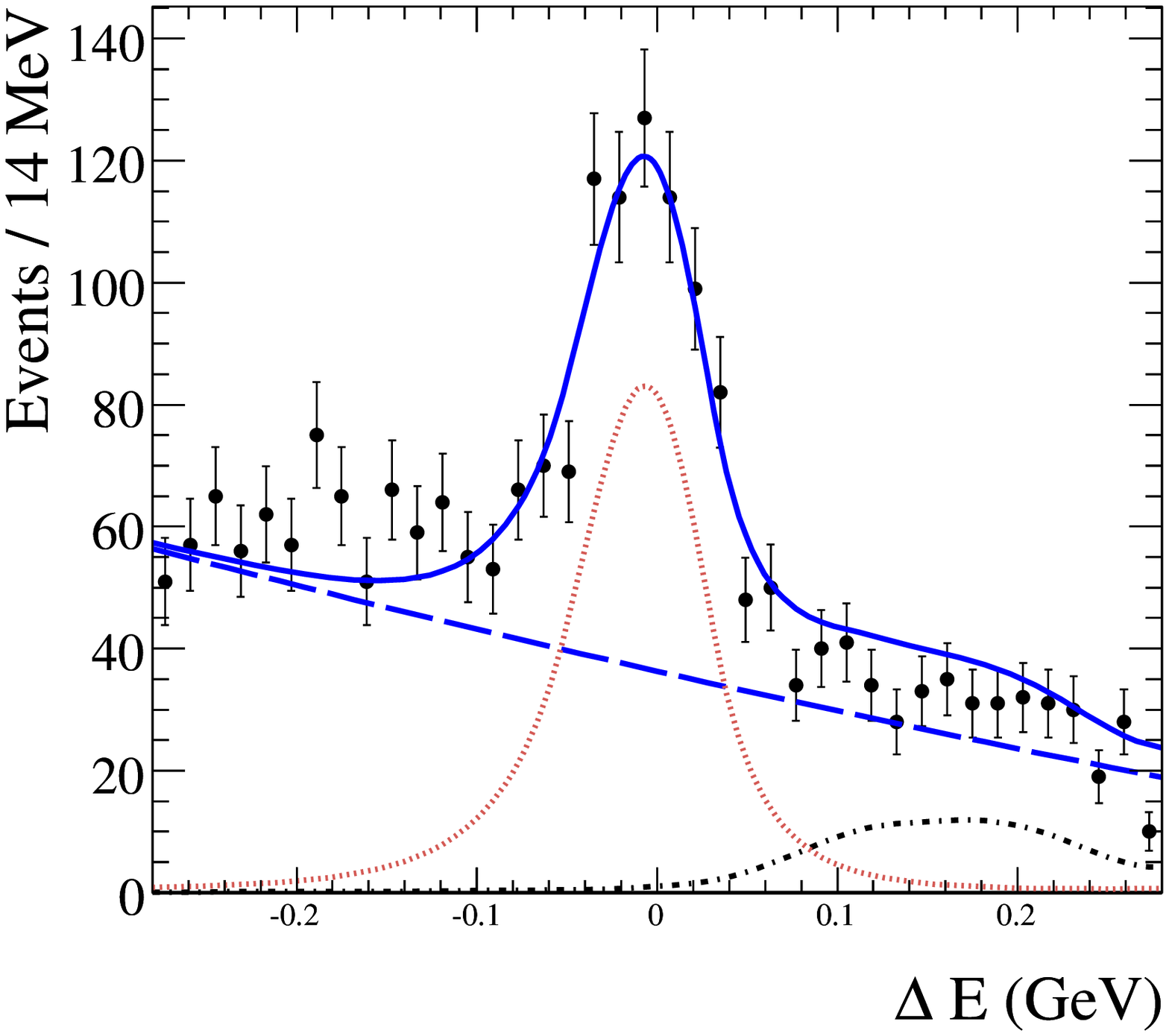}
\includegraphics[width=0.300\linewidth]{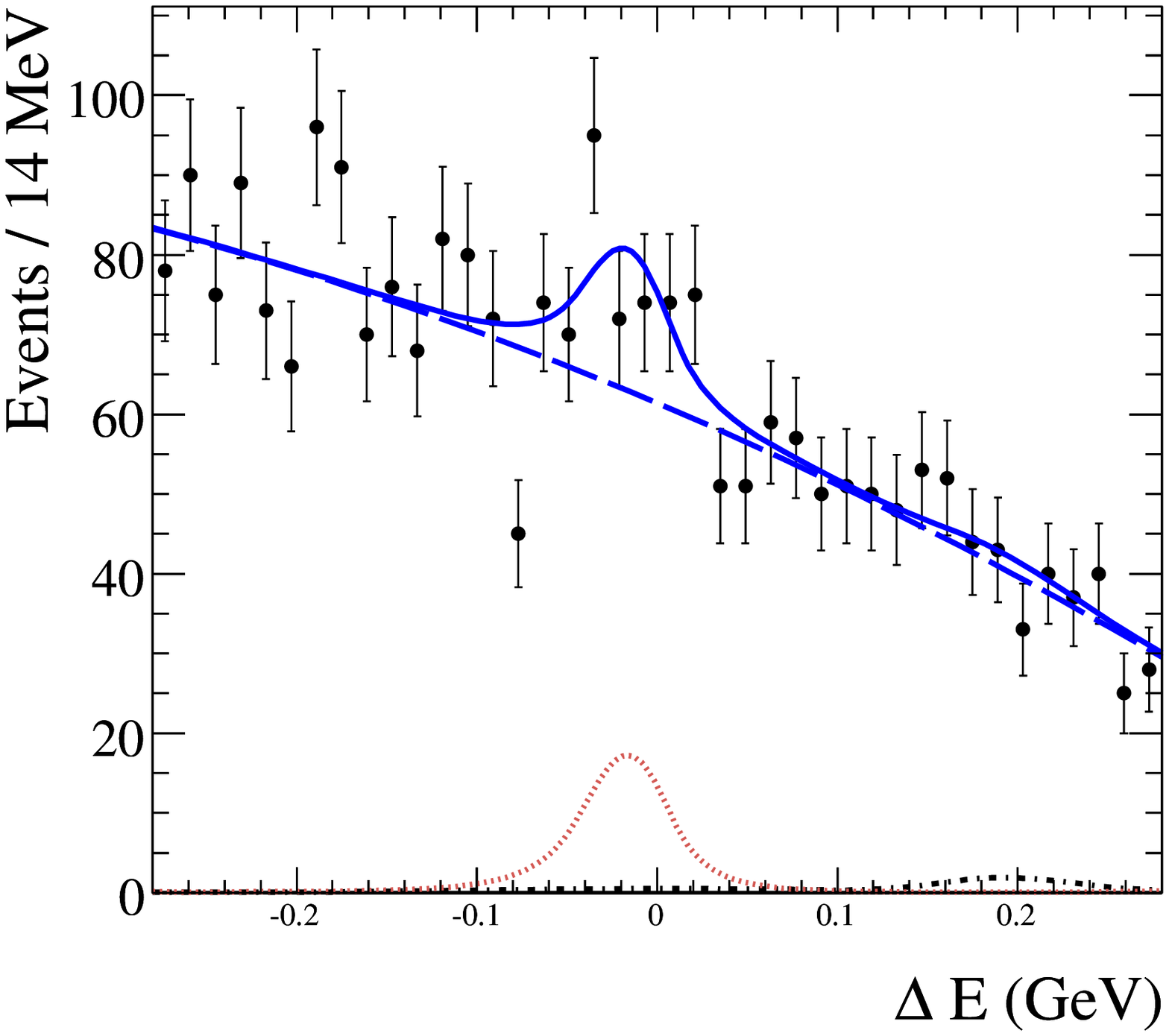}
\includegraphics[width=0.300\linewidth]{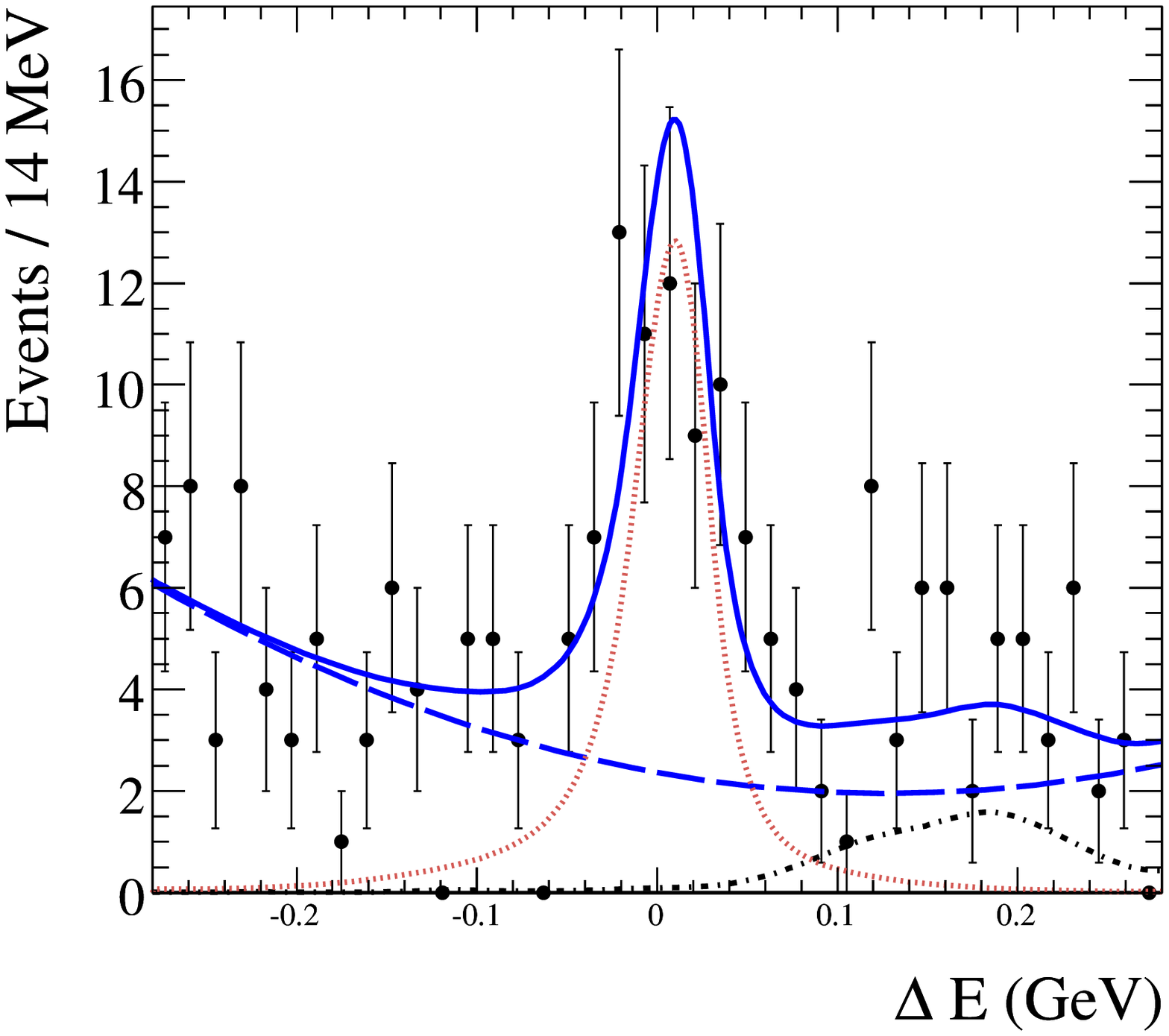}
\caption{\label{fig:fitData_2} Fit of $\DeltaE$ distributions in
data for modes $\Bzb\ra\Dstarz\piz$ (top left),
$\Bzb\ra\Dstarz\omega$ (top middle),
$\Bzb\ra\Dstarz\eta(\pi\pi\piz)$ (top right),
$\Bzb\ra\Dstarz\eta(\gg)$ (bottom left),
$\Bzb\ra\Dstarz(\Dz\piz)\etapr(\rho^0\g)$ (bottom middle) and
$\Bzb\ra\Dstarz\etapr(\pi\pi\eta)$ (bottom right). Detailed legend
is provided in Figure \ref{fig:fitData_1}.}
\end{center}
\end{figure*}

The iterative fit procedure is applied to data. The fitted
$\DeltaE$ distributions for the sum of $\Dstze$ submodes are given
in Figures \ref{fig:fitData_1} and \ref{fig:fitData_2}.

The signal and background yields obtained from the fit to data are
given in Table~\ref{tab:nbEvtFitDATA} with the corresponding statistical
significances.

\begin{table*}[htb]
{\footnotesize \caption{\label{tab:nbEvtFitDATA} We give the
number of signal events ($S$), crossfeed ($N_{\textrm{cf}}$),
$\Bm\ra\Dstze\rho^-$ ($N_{\textrm{D}\rho}$) and combinatorial
background ($N_{\textrm{combi}}$) from the $\DeltaE$ fit of data,
and the $\BF$ measured for each submode. The number of background
events $N_{\textrm{cf}}$, $N_{\textrm{D}\rho}$, and
$N_{\textrm{combi}}$ are computed from the fitted pdf in the
window $|\DeltaE| < 100~\mev$. The {\it statistical significance}
is defined as $\sqrt{2ln(L_{S+B}/L_B)}$, where $L_{S+B}$ is the
likelihood of the fit in data for the hypothesis of a signal,
while $L_B$ is the likelihood for the hypothesis of pure background
only.}
\begin{center}
\begin{tabular}{lccccccc}
\hline\hline
$\Bzb$ mode&  $S$ & $N_{\textrm{cf}}$ & $N_{\textrm{D}\rho}$ & $N_{\textrm{combi}}$  & $\BF(\times 10^{-4})$ & Statistical  \\[3pt]
(decay channel) & & &  &  & $\pm\textrm{stat.}\pm\textrm{syst.}$ & significance \\[3pt]
 \hline
$\Dz\piz$ & 3369 $\pm$  102 & 135  &  440 $\pm$ 11 & 2585 $\pm$ 63 & 2.78  $\pm$  0.08  $\pm$ 0.20 &  35.5\\
$\Dz\eta(\gg)$ & 1054 $\pm$  49 & 44  &  - & 823 $\pm$ 21   &  2.34  $\pm$  0.11 $\pm$ 0.17& 26.1 \\
$\Dz\eta(\pi\pi\piz)$ & 454 $\pm$  29 & 20  &  - & 487 $\pm$ 15   &  2.51  $\pm$  0.16 $\pm$ 0.17& 20.3\\
$\Dz\omega$ & 1400 $\pm$  58 & 68  &  - & 1648 $\pm$ 28  &  2.77  $\pm$  0.13 $\pm$ 0.22&  29.4\\
$\Dz\etapr(\pi\pi\eta(\gg))$ & 134 $\pm$  14 & 6  &  - & 74 $\pm$ 6   & 1.29  $\pm$  0.14 $\pm$ 0.09& 14.7\\
$\Dz\etapr(\rho^0\gamma)$ & 290 $\pm$  42 & 13  &  - & 2382 $\pm$ 33  &  1.95  $\pm$  0.29 $\pm$ 0.30 & 7.2\\
$\Dstarz\piz$ & 958 $\pm$  73 & 121  &  1218 $\pm$ 41 & 844 $\pm$ 56 &   1.78  $\pm$  0.13 $\pm$ 0.23 & 15.1\\
$\Dstarz\eta(\gg)$ & 629 $\pm$  39 & 33  &  - & 525 $\pm$ 18 &   2.37  $\pm$  0.15 $\pm$ 0.24 & 19.4 \\
$\Dstarz\eta(\pi\pi\piz)$ & 241 $\pm$  25 & 11  &  - & 341 $\pm$ 13  &   2.27  $\pm$  0.23 $\pm$ 0.18& 12.2\\
$\Dstarz\omega$ & 1692 $\pm$  86 & 60  &  - & 4507 $\pm$ 47 &   4.44  $\pm$  0.23 $\pm$ 0.61 & 22.3\\
$\Dstarz\etapr(\pi\pi\eta)$ & 61 $\pm$  10 & 3  &  - & 34 $\pm$ 4  &  1.12  $\pm$  0.26 $\pm$ 0.27 & 8.0\\
$\Dstarz(\Dz\piz)\etapr(\rho^0\gamma)$ & 86 $\pm$  28 & 6  &  - & 874 $\pm$ 20  &   1.64  $\pm$  0.53 $\pm$ 0.20 & 3.3\\
\hline\hline
\end{tabular}
\end{center}
}
\end{table*}

\section{SYSTEMATIC STUDIES}
\label{sec:Systematics}

There are several sources of systematic uncertainty in this
analysis. Table~\ref{tab:SystCombi3} summarizes the studied systematic uncertainties with their values in percentage of
$\BF$ for the different modes $\Dstze\hz$. The large uncertainties for $\Dstarz\hz$ mainly come from $\Dstarz(\Dz\g)\hz$.

The categories ``$\piz/\gamma$ detection'' (``Tracking'') account for the systematics on the
reconstruction of $\piz/\g$ and charged particle
tracks, and are taken as the uncertainty on the efficiency
correction computed in the study of $\tau$ decays (see
Section~\ref{sec:MCcorrection}).

Similarly, the systematic uncertainties on kaon identification and
$\KS$ selections, are estimated from the
uncertainties on MC efficiency corrections computed in the study
of pure samples of kaons and $\KS$ mesons in data respectively
(see Section~\ref{sec:MCcorrection}).

The systematic errors on the submodes $\BF$ is a combination of the uncertainty
on each $\Dz$ and $\hz$ submode~\cite{ref:PDG}. The correlation between the calibration mode
$\Dz\ra\Km\pip$ and $\Dz\ra X$, $X\ne\Km\pip$ was accounted for.

The uncertainty related to the number of $\BB$ pairs and to the
limited MC statistics is also included~\cite{ref:Babar2004}.

The systematics on resonance mass cut are computed as the relative difference of
signal yield when the mass mean and resolution for the selection are
taken from a fit in data.

The uncertainties for the $\qqbar$ rejection and the $\Dstze$
selections are obtained from the study performed on the control
sample $\Bm\ra\Dstze\pim$ and are estimated as the uncertainty on
the efficiency correction eff(data)$/$eff(MC), including the correlations between the samples before and after selections (see
Section~\ref{sec:MCcorrection}).

The uncertainties for the cuts on $\rho^0$ and $\Dz\omega$ helicities are obtained by varying the selection cut
values by $\pm 10~\%$ around the maximum of statistical
significance.

The category ``Signal shape'' represents the uncertainty on the
shape of the signal pdf for the $\DeltaE$ fit. The expected shape
difference between data and MC is estimated from a study of the
control sample $\Bm\ra\Dstze\rho^-$, hich gives:
$|\Delta\textrm{mean}| \simeq 5.7~\mev$ and $|\Delta\sigma| \simeq
2.3~\mev$. The systematic on signal shape is obtained by varying
the pdf parameters by $\pm 5.7~\mev$ for the mean and $\pm
2.3~\mev$ for the resolution.

The uncertainty on the continuum background shape is estimated
from the difference of the pdf fitted on generic MC's with the pdf
fitted in the $\mes$ sideband $5.24<\mes<5.26~\gevcc$ in data.
When a Gaussian is added to the combinatorial background shape,
the related uncertainty is computed by varying its means and resolution
by $\pm 1\sigma$.

The uncertainties related to the branching fractions of the
crossfeed from $\Dstze\hz$ and $\Dstze\rho^-$ were computed by varying the $\BF$ separately by $\pm1\sigma$ in the fit procedure.
The value of $\sigma$ used is taken from the fit for $\Dstze\hz$
and from PDG for $\Dstze\rho^-$.

The uncertainty related to the crossfeed shape, modelled by a
non-parametric pdf, is obtained by shifting and smearing the pdf mean and resolution
by $\pm5.7$ and 2.3~$\mev$ respectively, with a convolution by a Gaussian.

The uncertainty in ``$\Dstarz\omega$ helicity'' accounts for the
hypothesis on the $\Dstarz\omega$ polarization in the MC
generation. As that polarization had never been measured,
following HQET and factorization based arguments, the
$\Dstarz\omega$ polarization was simulated using the helicity
parameters measured in $\Bm\ra D^{*0}\rho^-$, thus a longitudinal
fraction of $f_L=0.869$~\cite{ref:CLEOhelicity}, confirmed by a
CLEO measurement~\cite{ref:CLEOhelicity2}. That
systematic is estimated by a comparison of the efficiency as obtained from an exclusive simulation of
$\Bzb\ra\Dstarz\omega$ with a low $f_L$ value of $0.08$ and is taken as half of
the relative difference of the signal yield.

The most significant systematic uncertainties come from the $\piz/\g$
reconstruction, the submodes $\BF$, the longitudinal fraction for
$\Dstarz\omega$, from the signal and background parametrization on
the $\DeltaE$ fit and from the crossfeed by $\Dstze\rho^-$.

\begin{table*}[htb]
{\footnotesize
\caption{\label{tab:SystCombi3}Systematic uncertainties ($\%$) for
the modes $\Bzb\ra\Dstze\hz$ where $\Dstze$ submodes are summed.}
\begin{center}
\begin{tabular}{l|cccccc|cccccc}\hline\hline
 & \multicolumn{6}{|c|}{$\Dz\hz$} & \multicolumn{6}{c}{$\Dstarz\hz$} \\
Sources & $\piz$ & $\eta(\gg)$ & $\eta(3\pi)$ &
$\omega$ & $\etapr(\pi\pi\eta)$ & $\etapr(\rho^0\g)$  & $\piz$ & $\eta(\gg)$ & $\eta(3\pi)$
&$\omega$ & $\etapr(\pi\pi\eta)$ & $\etapr(\rho^0\g)$ \\
\hline
$\piz/\gamma$ detection& 4.0& 4.0&  4.0 & 4.0 & 4.0  & 2.4 & 5.9 & 5.9&  6.1 & 6.1 & 5.8  & 4.5 \\
PID & 1.1& 1.1&  1.2 & 1.2 & 1.2  & 1.1 & 1.2 & 1.2&  1.2 & 1.2 & 1.2  & 1.2 \\
Tracking & 0.9& 0.9&  1.6 & 1.6 & 1.6  & 1.6 &  0.9& 0.9&  1.6 & 1.6 & 1.5& 1.5 \\
$\KS$ selection & 0.1& 0.1&  0.1 & 0.1 & 0.1 & 0.1 & 0.0& 0.1&  0.0 & 0.1 & -  & - \\
Submodes $\BF$ & 3.2& 3.3&  3.5 & 3.3 & 4.5& 4.4  & 3.1 & 3.2 & 3.4 & 3.2  & 4.3 & 6.4 \\
$\BB$ counting & 1.1& 1.1&  1.1 & 1.1 & 1.1 & 1.1 & 1.1& 1.1&  1.1 & 1.1 & 1.1  & 1.1 \\
$MC$ statistics & 0.3& 0.5&  0.7 & 0.4 & 1.0 & 0.9  & 0.6& 0.8&  1.1 & 0.6 & 1.1 & 1.2 \\
Resonances mass & 0.5& 0.7&  0.4 & 0.1 & 0.7  & 0.7 & 0.5 & 0.7  & 0.4  & 0.1   & 0.7   & 0.7 \\
$\qqbar$ rejection & 0.2& 0.2&  0.2 & 0.2 & 0.2  & 0.2 & 0.2& 0.2&  0.2 & 0.2 & 0.2 & 0.2 \\
$\Dstze$ selection  & 0.3& 0.3&  0.3 & 0.3 & 0.3  & 0.3 & 0.4& 0.4&  0.4 & 0.4 & 0.4  & 0.4 \\
$\rho^0$ helicity  & -& -&  - & - & -  & 2.1 & -& -&  - & - & -  & 2.1 \\
$\Dstze\omega$ helicity & -& -&  - & 1.9 & -  & - & -& -&  - & 10.8 & -  & - \\
Signal shape & 2.6& 2.8&  1.5 & 1.7 & 1.3  & 2.7 & 1.9& 2.8&  1.8 & 3.3 & 1.0  & 3.1 \\
Continuum shape & 1.0 & 0.8&  1.1 & 4.2 & 1.2  & 13.3 & 0.9 & 6.5&  1.7 & 3.3 & 21.2  & 6.8 \\
Crossfeed $\BF$ & 0.0& 0.0&  0.0 & 0.0 & 0.1  & 0.1 & 8.6 & 1.2&  2.11 & 0.8 & 8.6 & 0.3 \\
Crossfeed shape & 0.7& 0.7&  0.3 & 0.6 & 0.5 & 0.4 &  6.1& 0.2&  0.3 & 0.2 & 0.6 & 0.2 \\\hline
Total & 6.2& 6.3&  6.2 & 7.5 & 6.8 & 14.9  &  12.8 & 10.0&  8.1 & 13.8 & 24.1  & 11.3 \\
\hline\hline
\end{tabular}
\end{center}
}
\end{table*}

\section{RESULTS}\label{se:results}

\subsection{$\BF$ measurement}

The branching fractions obtained from the $\DeltaE$ fit on data
are given in Table~\ref{tab:nbEvtFitDATA} for the sum of the
$\Dstze$ submodes. The consistency with $\Dz$ submodes was
checked. All the quoted results are the branching
fraction $\BF_{\textrm{sum}}$ computed with the fit of the $\Dz$
sum.

The statistical uncertainty $\sigma_{\textrm{stat}}$ on
$\BF_{\textrm{sum}}$ is provided by the fit while the systematic
uncertainty is computed as the average of the systematic weighted
by the total corrected acceptance $\mathcal{A}_i$:
\begin{equation}\label{systUnc}
\sigma_{\textrm{syst}} = \frac{\sum_i
\mathcal{A}_i\cdot\sigma_{\textrm{syst,}i}}{\sum_i \mathcal{A}_i},
\end{equation}
where
$(\BF_i\pm\sigma_{\textrm{syst,}i}\pm\sigma_{\textrm{stat,}i})$ is
the measured branching fraction with the submode $i=\Km\pip$,
$\Km\pip\pim\pip$, $\Km\pip\piz$ or $\KS\pip\pim$. The total
uncertainty is eventually computed as the quadrature sum of
$\sigma_{\textrm{stat}}$ and $\sigma_{\textrm{syst}}$. We checked
the compatibility of $\BF_{\textrm{sum}}$ with the measurements in
the submodes $\BF_i$ for significant signal. The $\BF$'s measured
with $\Dstarz$ submodes are compatible with $\BF_{\textrm{sum}}$,
except for $\Bzb\ra\Dstarz\piz$ where the incompatibility between
$\Dstarz\ra\Dz\g$ and $\Dstarz\ra\Dz\piz$ is due to the large
background contribution from $\Dstze\rho^-$. Likewise the
measurements in the different $\eta$ or $\etapr$ submodes are
compatible. The measurements in the $\eta^{(')}$ submodes are
combined through a $\BF$ average weighted by the total uncertainty
on $\BF$:

\begin{equation}
\BF_{\textrm{avg}} = \frac{\sum_i \BF_i/\sigma_i^2}{\sum_i
1/\sigma_i^2},
\end{equation}
where $(\BF_i\pm\sigma_i)$ is the measured branching fraction with
the submode $i$, and $\sigma_i$ the total uncertainty on $\BF_i$.
The statistical uncertainty on the $\BF$ combination is :

\begin{equation}
\sigma^2_{\textrm{stat,avg}} =
\sum_i\frac{1}{1/\sigma_{\textrm{stat,}i}^2},
\end{equation}
and the systematic is obtained from the Equation (\ref{systUnc}).

The $\BF$'s measured in this analysis are compatible with the
previous measurements by $\babar$~\cite{ref:Babar2004},
CLEO~\cite{ref:Cleo2002} and
Belle~\cite{ref:Belle2005,ref:Belle2006}. For the modes
$\Bzb\ra\Dstze\eta$, $\Dstarz\omega$ and $\Dstarz(\Dz\piz)\piz$,

our measured $\BF$s are significantly higher than the ones measured by Belle~\cite{ref:Belle2006}. We also measured the $\BF$
in the dataset studied previously by
$\babar$~\cite{ref:Babar2004} and found compatible $\BF$ values
with systematic uncertainties lowered by about 25 to 50~$\%$. The improvement mainly comes from the selections on $\Dstze$ and on the $\qqbar$ background rejection.

\subsection{Isospin analysis}

The isospin $I$ symmetry relates the amplitudes of the decays
$\Bm\ra D^{(*)0}\pim$, $\Bzb\ra D^{(*)+}\pim$ and $\Bzb\ra
D^{(*)0}\piz$, which can be written as linear combinations of the
isospin eigenstates $\mathcal{A}_I$, $I=1/2$,
$3/2$~\cite{ref:NeubertPetrov,ref:rosner}:

\begin{eqnarray}
&\mathcal{A}(\Dstze\pim) =&
\sqrt{3}\mathcal{A}_{3/2,D^{(*)}},\nonumber\\
&\mathcal{A}(D^{(*)+}\pim) =& 1/\sqrt{3}\mathcal{A}_{3/2,D^{(*)}}
+
\sqrt{2/3}\mathcal{A}_{1/2,D^{(*)}},\nonumber\\
&\mathcal{A}(D^{(*)0}\piz) =& \sqrt{2/3}\mathcal{A}_{3/2,D^{(*)}}
- \sqrt{1/3}\mathcal{A}_{1/2,D^{(*)}},
\end{eqnarray}
and so:

\begin{equation}
\mathcal{A}(\Dstze\pim) = \mathcal{A}(D^{(*)+}\pim) +
\sqrt{2}\mathcal{A}(D^{(*)0}\piz).
\end{equation}

The relative strong phase between the eigenstates
$\mathcal{A}_{1/2}$ and $\mathcal{A}_{3/2}$ is noted as $\delta$ for the
$D\pi$ system and $\delta^*$ for $D^*\pi$ system. Final State
Interactions between the states $\Dstze\piz$ and $D^{(*)+}\pim$
may lead to a value of $\delta^{(*)}$ different from zero and,
through constructive interference, to a higher value of $\BF$ for
$\Dstze\piz$. One can also define the amplitude ratio $R^{(*)}$:

\begin{equation}
R^{(*)}=\frac{|\mathcal{A}^{(*)}_{1/2}|}{\sqrt{2}|\mathcal{A}^{(*)}_{3/2}|}.
\end{equation}

In the HQET limit~\cite{ref:BBNS,ref:ChengYang}, the factorization model predicts
$\delta^{(*)}=0$ and
$R^{(*)}=1$, while SCET~\cite{ref:SCET_1,ref:SCET_2,ref:SCET_3}
predicts $\delta$ to have the same value in the $D\pi$ and $D^*\pi$
systems. The
strong phase $\delta^{(*)}$ can be computed with an isospin
analysis of the $D^{(*)}\pi$ system. The input parameters are the
world average values provided by the PDG 2007 partial update~\cite{ref:PDG}
of $\BF(\Bm\ra D^{(*)0}\pim)$, $\BF(\Bzb\ra D^{(*)+}\pim)$ and of
the $B$ lifetime ratio $\tau(\Bp)/\tau(\Bz)$, the value of
$\BF(\Bzb\ra\Dstze\piz)$ is the one measured in this analysis. We
calculate the value of $\delta^{(*)}$ and $R^{(*)}$ using a
frequentist approach~\cite{ref:ckmfitter}
$\delta=(30.2^{+3.2}_{-4.5})^{\circ}$ and $R=0.70^{+0.05}_{-0.05}$ for
$D\pi$ final states, and $\delta^{*}=(13.6^{+9.9}_{-13.6})^{\circ}$,
$R^*=0.64^{+0.06}_{-0.05}$ for $D^*\pi$ final states.

In both $D^*\pi$ and $D\pi$ cases, the amplitude ratio is
significantly different from the factorization prediction
$R^{(*)}=1$. The strong phase is significantly different from zero
in the system $D\pi$, which points out that non-factorizable FSI
are not negligible. In the system $D^*\pi$ that phase $\delta^*$
is smaller and compatible with zero due to the lower value of the
$\BF$ for $\Dstarz\piz$.

\subsection{Comparison to theoretical predictions on $\BF$'s}

Table~\ref{tab:comparaisonTheorie_1} compares the $\BF$
measured with this analysis to the predictions by factorization~\cite{ref:Chua,ref:Neubert,ref:Deandrea,ref:Deandrea2}
and pQCD~\cite{ref:pQCD_1,ref:pQCD_2}. We confirm the conclusion by the previous $\babar$
analysis, the values measured being higher by a factor of about 4 than the values predicted by factorization. The pQCD predictions are
closer to experimental values but globally higher except for $\Dz\piz$.

\begin{table*}[htb]
{\footnotesize \caption{\label{tab:comparaisonTheorie_1}Comparison of the measured branching fraction $\BF$ in units of $10^{-4}$, with the predictions by factorization and pQCD~\cite{ref:pQCD_1,ref:pQCD_2}. The first quoted uncertainty is
 statistical and the second is systematic.}
\begin{center}
\begin{tabular}{lccc}
\hline\hline
$\Bzb$ mode & This measurement $(\times 10^{-4})$  & factorization~$(\times 10^{-4})$ & pQCD~$(\times 10^{-4})$ \\
\hline
  $\Dz\piz$ &   2.78  $\pm$  0.08 $\pm$ 0.20 & 0.58~\cite{ref:Chua} 0.70~\cite{ref:Neubert} & 2.3-2.6 \\
  $\Dstarz\piz$ &   1.78  $\pm$  0.13$\pm$ 0.23 &   0.65~\cite{ref:Chua} 1.00~\cite{ref:Neubert}& 2.7-2.9 \\
  $\Dz\eta$ &   2.41  $\pm$  0.09 $\pm$ 0.17   &   0.34~\cite{ref:Chua} 0.50~\cite{ref:Neubert} & 2.4-3.2\\
  $\Dstarz\eta$ &  2.32  $\pm$  0.13 $\pm$ 0.22 &   0.60~\cite{ref:Neubert} & 2.8-3.8 \\
  $\Dz\omega$ &   2.77  $\pm$  0.13 $\pm$ 0.22 &   0.66~\cite{ref:Chua} 0.70~\cite{ref:Neubert} & 5.0-5.6 \\
  $\Dstarz\omega$ &   4.44  $\pm$  0.23 $\pm$ 0.61 &   1.70~\cite{ref:Neubert}  & 4.9-5.8  \\
 $\Dz\etapr$ &   1.38  $\pm$  0.12 $\pm$ 0.22 &   0.30-0.32~\cite{ref:Deandrea2}; 1.70-3.30~\cite{ref:Deandrea} & 1.7-2.6  \\
 $\Dstarz\etapr$ &   1.29  $\pm$  0.23 $\pm$ 0.23   &   0.41-0.47~\cite{ref:Deandrea}  & 2.0-3.2  \\
\hline\hline
\end{tabular}
\end{center}
}
\end{table*}

Factorization predicts the ratio
$\BF(\Bzb\ra\Dstze\etapr)/\BF(\Bzb\ra\Dstze\eta)$ to have a value between
0.64 and 0.68~\cite{ref:Deandrea} linked to the $\eta-\etapr$
mixing. Table \ref{tab:comparaisonTheorie_2} compares that
prediction with the experimental measurements. The measured ratios
are smaller than the prediction and are compatible at 1~$\sigma$.\\

The effective theory
SCET~\cite{ref:SCET_1,ref:SCET_2,ref:SCET_3} does not
predict the absolute value of the $\BF$ but that the ratios
$\BF(\Bzb\ra\Dstarz\hz)/\BF(\Bzb\ra\Dz\hz)\sim1$ for $\hz =
\piz,\eta$ and $\etapr$. For $\hz=\omega$ that prediction holds
only for the longitudinal component of $\Bzb\ra\Dstarz\omega$ as
long-distance QCD interactions may increase the transverse
amplitude. A measurement of $f_L$ is therefor needed.

The SCET gives also a prediction about the ratio
$\BF(\Bzb\ra\Dstze\etapr)/\BF(\Bzb\ra\Dstze\eta)\simeq 0.67$,
which is similar to the prediction by factorization.

The $\BF$ ratios are given in
Table~\ref{tab:comparaisonTheorie_2}, the common systematics
between $\Dz\hz$ and $\Dstarz\hz$ are cancelled. The ratios
$\BF(\Bzb\ra\Dstarz\hz)/\BF(\Bzb\ra\Dz\hz)$ for
$\hz=\piz,\eta,\etapr$ are compatible with 1.

\begin{table}[htb]
{\footnotesize \caption{\label{tab:comparaisonTheorie_2}Ratios
of branching fractions $\BF(\Bzb\ra\Dstarz\hz)/\BF(\Bzb\ra\Dz\hz)$. The first uncertainty is statistical, the second is systematic.}
\begin{center}
\begin{spacing}{2}
\begin{tabular}{lc}
\hline\hline $\BF$ ratio & This measurement\\\hline
$\frac{\BF(\Dstarz\piz)}{\BF(\Dz\piz)}$ & 0.64 $\pm$ 0.05
$\pm$ 0.08 \\
$\frac{\BF(\Dstarz\eta(\gg))}{\BF(\Dz\eta(\gg))}$ & 1.01
$\pm$ 0.08 $\pm$ 0.08 \\
$\frac{\BF(\Dstarz\eta(\pi\pi\piz))}{\BF(\Dz\eta(\pi\pi\piz))}$ &
0.91
$\pm$ 0.11 $\pm$ 0.04 \\
$\frac{\BF(\Dstarz\eta)}{\BF(\Dz\eta)}$ & 0.96
$\pm$ 0.06 $\pm$ 0.07 \\
$\frac{\BF(\Dstarz\omega)}{\BF(\Dz\omega)}$ & 1.61
$\pm$ 0.11 $\pm$ 0.20 \\
$\frac{\BF(\Dstarz\etapr(\pi\pi\eta))}{\BF(\Dz\etapr(\pi\pi\eta))}$
& 0.87
$\pm$ 0.22 $\pm$ 0.04 \\
$\frac{\BF(\Dstarz\etapr(\rho^0\g))}{\BF(\Dz\etapr(\rho^0\g)))}$ &
0.84
$\pm$ 0.30 $\pm$ 0.15 \\
$\frac{\BF(\Dstarz\etapr)}{\BF(\Dz\etapr))}$ & 0.93
$\pm$ 0.19 $\pm$ 0.11 \\
$\frac{\BF(\Dz\etapr)}{\BF(\Dz\eta)}$ & 0.57 $\pm$ 0.06
$\pm$ 0.06 \\
$\frac{\BF(\Dstarz\etapr)}{\BF(\Dstarz\eta)}$ & 0.55 $\pm$ 0.11
$\pm$ 0.04 \\
\hline\hline
\end{tabular}
\end{spacing}
\end{center}
}
\end{table}

\section{CONCLUSIONS}
\label{sec:Conclusions} We measured the branching fractions of the
color-suppressed decays $\Bzb\ra\Dstze\hz$, where $\hz=\piz$, $\eta$,
$\omega$, and $\etapr$ with 454$\times 10^6\BB$ pairs. Our measurements are
in agreement with the previous
results~\cite{ref:Cleo2002,ref:Babar2004,ref:Belle2005}
with a significant decrease of both statistical and systematic
uncertainties. We confirm the significant difference from
theoretical predictions by factorization quoted in the previous
analysis~\cite{ref:Babar2004} and provide strong constraints on
the models of color-suppressed decays.

\section{ACKNOWLEDGMENTS}
\label{sec:Acknowledgments}


\input acknow_PRL

\end{document}

%% file: authors_ICHEP2008.tex
\begin{center}
\small

The \babar\ Collaboration,
\bigskip

%
B.~Aubert,
M.~Bona,
Y.~Karyotakis,
J.~P.~Lees,
V.~Poireau,
E.~Prencipe,
X.~Prudent,
V.~Tisserand
\inst{Laboratoire de Physique des Particules, IN2P3/CNRS et Universit\'e de Savoie, F-74941 Annecy-Le-Vieux, France }
J.~Garra~Tico,
E.~Grauges
\inst{Universitat de Barcelona, Facultat de Fisica, Departament ECM, E-08028 Barcelona, Spain }
% ...

\end{center}\newpage

%% file: acknow_PRL.tex
We are grateful for the excellent luminosity and machine conditions
provided by our \pep2\ colleagues, 
and for the substantial dedicated effort from
the computing organizations that support \babar.
The collaborating institutions wish to thank 
SLAC for its support and kind hospitality. 
This work is supported by
DOE
and NSF (USA),
NSERC (Canada),
CEA and
CNRS-IN2P3
(France),
BMBF and DFG
(Germany),
INFN (Italy),
FOM (The Netherlands),
NFR (Norway),
MES (Russia),
MEC (Spain), and
STFC (United Kingdom). 
Individuals have received support from the
Marie Curie EIF (European Union) and
the A.~P.~Sloan Foundation.

%% file: conf_sum2008.bbl
\begin{thebibliography}{99}
\bibitem{ref:bukin}
{\footnotesize The modified Novossibirsk function divides the
fitting region in a peaking region, a low and a high tail
regions. For a variable $x$, the modified Novosibirsk function
$f(x)$ is:

in the peak region : $x_1 < x < x_2$,
\begin{equation}
f(x) = A_p \times \exp{\left( -\ln(2)\left(
    \frac{\ln\left(1+2\tau\sqrt{\tau^2+1}\cdot\frac{x-x_p}{\sigma_p\sqrt{2\ln(2)}}
\right)}{\ln(1+2\tau^2-2\tau\sqrt{\tau^2+1})} \right)^2 \right)},
\end{equation}
in the low tail region: $x<x_1$,
\begin{equation}
f(x) = A_p \times \exp{\left(
\frac{\tau\sqrt{\tau^2+1}\cdot(x-x_1)
\sqrt{2\ln(2)}}{\sigma_p(\sqrt{\tau^2+1}-\tau)^2\ln(\sqrt{\tau^2+1}+\tau)}
+ \rho_1\cdot\left(\frac{x-x_1}{x_p-x_1}\right)^2 - \ln2 \right)}.
\end{equation}
in the high tail region: $x>x_2$,
\begin{equation}
f(x) = A_p \times \exp{\left(-
\frac{\tau\sqrt{\tau^2+1}\cdot(x-x_2)
\sqrt{2\ln(2)}}{\sigma_p(\sqrt{\tau^2+1}+\tau)^2\ln(\sqrt{\tau^2+1}+\tau)}
+ \rho_2\cdot\left(\frac{x-x_2}{x_p-x_2}\right)^2 - \ln2 \right)}.
\end{equation}
The parameters are:

\begin{itemize}
\item $A_p$ is the value at the maximum of the function,

\item $x_p$ is the peak position,

\item $\sigma_p$ is the width of the peak defined as the width at
half-height divided by 2.35.

\item $\xi$ is an asymmetry parameter.

\end{itemize}
The positions $x_1$ and $x_2$ are $x_{1,2} = x_p + \sigma_p\sqrt{2\ln 2}\cdot\left(
\frac{\xi}{\sqrt{\xi^2+1}}\mp 1 \right)$,

}
\end{thebibliography}
